\documentclass[revtex4,aps,nofootinbib,showpacs]{revtex4}

\usepackage{amsmath,amsfonts,amssymb,amsthm}           
\usepackage{dsfont}                                    
\usepackage{hyperref}                                  
\usepackage{graphicx, epsfig}                          

\newcommand{\sDelta}{{\scriptstyle\Delta}}
\newcommand{\idMatrix}{\mathds{1}}
\newcommand{\pd}{\partial}
\newcommand{\ii}{\mathrm{i}}
\newcommand{\const}{\mathrm{const}}
\newcommand{\diff}{\mathrm{d}}

\newcommand{\bigO}{\mathit{O}}           
\DeclareMathOperator{\sgn}{sgn}          
\DeclareMathOperator{\TProd}{\mathrm{T}} 
\DeclareMathOperator{\tr}{Sp}            
\newcommand{\hc}{^\dagger}               
\newcommand{\avg}[1]{\langle{#1}\rangle} 
\newcommand{\degree}{^\circ}             

\begin{document}
\title{Analytical treatment of long-term observations of the day-night asymmetry for solar neutrinos}
\author{Sergey~S.~Aleshin} \email{aless2001@mail.ru}
\author{Oleg~G.~Kharlanov} \email{okharl@mail.ru}
\author{Andrey~E.~Lobanov} \email{lobanov@phys.msu.ru}
\affiliation{Department of Theoretical Physics, Faculty of Physics, Moscow State University, 119991
Moscow, Russia}

\begin{abstract}
   The Earth's density distribution can be approximately considered piecewise continuous at the
   scale of two-flavor oscillations of typical solar neutrinos, such as the beryllium-7 and
   boron-8 neutrinos. This quite general assumption appears to be enough to analytically calculate
   the day-night asymmetry factor for such neutrinos. Using the explicit time averaging procedure, we show that,
   within the leading-order approximation, this factor is determined by the
   electron density within about one oscillation length under the detector,
   namely, in the Earth's crust (and upper mantle for high-energy neutrinos).
   We also evaluate the effect of the inner Earth's structure on the observed asymmetry
   and show that it is suppressed and mainly comes from the neutrinos observed near the winter and summer
   solstices. As a result, we arrive at the strict interval constraint on the asymmetry,
   which is valid within quite a wide class of Earth models.
\end{abstract}

\maketitle


\section{Introduction}
The effect of neutrino oscillations in vacuum lies beyond the Standard Model and is thus
interesting both from the theoretical and experimental points of view. The oscillations in medium
have also been studied since Wolfenstein, who showed that neutrinos acquire a specific
flavor-dependent potential due to coherent forward scattering on the matter \cite{Wolfenstein}. As
a result, the neutrino propagation in medium should demonstrate the conversion from one flavor into
another (i.e. flavor oscillations), even if the vacuum mixing is negligible. This spectacular
result is known as the Mikheev--Smirnov--Wolfenstein effect \cite{MikheevSmirnov} and suffices to
explain the deficit of observed solar electron neutrinos \cite{Bethe}. According to Mikheev and
Smirnov, the leading-order estimation for the electron neutrino flux depends only on the points
where the neutrino was born (the core of the Sun) and absorbed (the detector). However, the
properties of the medium between these two points can also slightly affect the flavor composition
of the observed neutrino flux, leading, in particular, to the day-night (solar neutrino flavor)
asymmetry \cite{Carlson, Baltz}. The latter effect being crucial for the entire flavor oscillations
framework, a number of experiments were set up to catch this slight flavor composition variation
resulting from the nighttime neutrino propagation through the Earth \cite{SNO_DNA, SuperK_DNA,
Borexino}. Some experiments, which should be sensitive enough to distinguish the effect, are also
planned in the next decade and are now under construction \cite{LENA}.

However, one should hold in mind that, in order to make a conclusion on the expected
presence of the day-night asymmetry, one needs not only the experimental data (in terms of event
rates and energy spectra) but also a convenient theoretical prediction. First experiments in
neutrino oscillations were only aimed at outlining the domains in the neutrino mass and mixing
parameter space which do not contradict the observed rates. For such needs, the theoretical
estimations obtained using numerical simulations are quite acceptable (see, e.g.,
\cite{BahcallKrastev}). Indeed, although numerical experiments can lack perfect accuracy in some
regimes and do not give general results, the calculation of the desired effect can be performed for
the entire parameter space without drastic algorithm changes. Therefore, using numerical
predictions, one could more or less interpret the experimental data in terms of domains in this
parameter space which are excluded. But now that various neutrino experiments have
yielded a considerable amount of data on the vacuum neutrino mixing \cite{SNO, KamLand, Borexino1}
and its parameters are fixed quite firmly, the question is whether other neutrino
oscillation effects to be observed are consistent with the framework accepted so far. Namely, one
may ask: how should one compare the present and future experimental data on the day-night asymmetry
with the theory of neutrino oscillations in order to make an ultimate conclusion about their
agreement within the allowed class of Earth and solar models? The latter issue requires a formalism
able to give strict constraints (in a form of an interval with fixed boundaries) on the predicted
asymmetry within certain approximations, but valid within quite a wide class of the Earth and solar
models. Developing such a formalism constitutes the principal goal of the present paper. In
particular, using it, we readily find the constraints on the day-night asymmetry for
beryllium and boron neutrinos observed in such experiments as SNO, Super-Kamiokande, and Borexino
\cite{SNO_DNA, SuperK_DNA, Borexino}.

\vspace{1em}

Within the neutrino oscillations framework, it is common to use the Schroedinger-like equation to
describe the spatial variations of the neutrino flavor \cite{Wolfenstein, MikheevSmirnov}. Within
the two-flavor approximation, the Schroedinger problem is posed in terms of the $2\times 2$ flavor
evolution matrix (operator) $R(x,x_0)$, whose elements are the neutrino flavor transition
amplitudes after traveling from point $x_0$ to $x$. The evolution equation and the initial
condition read, respectively,
\begin{eqnarray}\label{eq1}
  \frac{\pd R(x,x_0)}{\pd x} &=& -\ii \lambda A(x) R(x,x_0),\\
  R(x_0,x_0) &=& \idMatrix,
\end{eqnarray}
where the matrix Hamiltonian is a point-dependent linear combination of the Pauli matrices,
\begin{gather}
 A(x) = a(x)\sigma_3 +b\sigma_1, \\
 a(x) = -\cos2\theta_0+\frac{2EV(x)}{\sDelta m^2}, \qquad  b = \sin2\theta_0. \label{eq5}
\end{gather}
Here, $\theta_0$ is the vacuum mixing angle, $E$ is the neutrino energy, $\sDelta m^2$ is the
difference between neutrino masses squared, and $V(x) = \sqrt{2} G_{\text{F}} N_e(x)$ is the
Wolfenstein potential, which is proportional to the electron density in the medium $N_e$ and the
Fermi constant $G_{\text{F}} = 1.17\times 10^{-11}\text{ MeV}^{-2}$. The constant coefficient $\lambda=\sDelta m^2/4E$ is the reciprocal
vacuum neutrino oscillation length, up to the factor of $\pi$,
\begin{equation}\label{eq:l_osc}
  \ell_{\mathrm{osc}}=\frac{4\pi E}{\sDelta m^2} = \frac{\pi}{\lambda}.
\end{equation}

Equation \eqref{eq1} defines a one-parametric subgroup of $SU(2)$ and hence of $SO(3)$, the
translation along $x$ being the group operation. In this sense, equation \eqref{eq1} is analogous
to the Bargmann--Michel--Telegdi equation \cite{b1} in the spinor representation \cite{b4}. It is
well known that the solution of the matrix linear ordinary differential equation, such as
\eqref{eq1}, can be represented as a time-ordered exponential (Dyson expansion)
\cite{DiffEquations}. However, in the case of general $N_e(x)$ profile, the solution in
terms of an analytical function of the matrix argument (i.e., without a symbolic operation such as
time ordering) appears to be too challenging to find. In the recent investigations, considerable
progress was made in finding exact solutions of Eq.~\eqref{eq1} in certain special cases
\cite{b5,b11}; nevertheless, the general approach to this kind of equation still remains to be
approximate.

It was first naturally accepted that numerical simulations could give exhaustive results here
and, in a sense, resolve the analytical difficulties that arise in the context of such
equations. For instance, a numerical approach was employed in \cite{BahcallKrastev,
Lisi} to find the domains in the neutrino mixing parameter space where the day-night
asymmetry should be potentially observable. However, as mentioned above, estimations
obtained with numerical techniques are not always reliable enough.
In particular, in the large-$\sDelta m^2$ regime, which has now proved to be realistic \cite{SNO,
KamLand}, the oscillation length \eqref{eq:l_osc} becomes small, and numerical evaluation of
rapidly oscillating solutions introduces large uncertainties. The numerical time averaging of the
flavor observation rates also becomes inaccurate in this regime. This effect is especially strong
for low-energy neutrinos, including beryllium neutrinos; probably, due to this fact, the
large-$\sDelta m^2$ and low-energy area was not covered by original numerical simulations
\cite{BahcallKrastev}. Therefore, obtaining stringent constraints on the day-night asymmetry
factor for the continuous measurement of small-length neutrino oscillations favors the analytical
approach.

Quite a large number of publications are devoted to the analysis of such approximate analytical
solutions. Probably the most effective technique for finding the approximate solutions of matrix
linear differential equations, such as \eqref{eq1}, is the so-called Magnus expansion
\cite{Magnus}, which is a generalization of the Baker--Campbell--Hausdorff formula \cite{Campbell}.
This approach provides the solution up to any order of approximation, as well as the constraints on
the remainder terms \cite{MagnusApps}. Unfortunately, this technique
\cite{Olivo90,Olivo,Olivo08,IoannisianSmirnov09}, as well as other general methods (see, e.g.,
\cite{Ohlsson, Akhmedov08}), does not readily  provide a way to find the solution in its explicit
and practically usable form, not firmly fixing the Earth model, namely, $N_e(x)$ density
distribution. It is thus desirable to find an approximate analytical solution of equation
\eqref{eq1}, which is valid under quite general assumptions about the electron density profile
$N_e(x)$. This idea was developed in \cite{Wei,IoannisianSmirnov04,AkhmedovValle}.

As it was mentioned earlier in this section, in our paper, we are not only aiming at finding the
relevant approximate expressions, but also at analyzing their accuracy and applicability domain.
Namely, in section~\ref{sec:EvolutionOperator}, we find the approximate solutions for the flavor
evolution matrix inside the Earth, and then, in section~\ref{sec:DNA}, we arrive at the observation
probabilities for the neutrinos of different flavors. These probabilities are finally subjected to
the averaging procedure due to the continuous observation of the solar neutrinos throughout the
year (Sec.~\ref{sec:Averaging}), and the results of this averaging are discussed in
section~\ref{sec:Discussion}. The magnitude of the day-night asymmetry appears to be sensitive to
the non-trivial structure of the Earth's crust under the neutrino detector, so the effect of the
crust is paid special attention in section~\ref{sec:CrustEffect}. The uncertainties of our
estimation are also discussed in section~\ref{sec:Discussion}, in the interesting cases of
beryllium and boron neutrinos. We analyze the sources of such uncertainties and finally compare our
results with the numerical simulation.  The detailed derivation of the approximate solutions for
the evolution operator is moved to the Appendix to make the flow of the paper less complicated.

\section{The Density Profile and the Evolution Matrix}\label{sec:EvolutionOperator}
In our investigation, we use the model electron density profile $N_e(x)$ with $n-1$ narrow
segments, where it changes steeply, separated by $n$ wide though sloping segments. Let us call
these segments cliffs and valleys, respectively. Let the cliffs be localized near points $x_{j}$,
$j=\overline{1,n-1}$, namely, occupy segments $[x_j^-, x_j^+]$ of widths $\epsilon_j$, where
$x_j^\pm \equiv x_j\pm\epsilon_j/2$. Then the valleys are $[x_{j-1}^+, x_{j}^-]$ and have the
widths $L_j \approx x_{j} - x_{j-1}$. In fact, this kind of model is a good approximation for the
Earth's density profile known in geophysics, where it corresponds to the so-called Preliminary
Reference Earth Model (PREM)\, \cite{Anderson,Anderson1}. In the present section, we will
assume that the above segments can be chosen in such a way that the inequality
\begin{equation}\label{piecewise_continuity}
\epsilon_j \ll \ell_{\text{osc}} \ll L_j
\end{equation}
takes place, i.e. the cliffs are narrow and the valleys are wide compared with the oscillation
length. In the following, we will briefly call such density distribution piecewise continuous.

Let us note in advance (see Sec.~\ref{sec:Discussion} for details) that for the energies
of beryllium neutrinos ($E = 0.862\text{MeV}$, $\ell_{\text{osc}} \approx 30~\text{km}$) and,
to some extent, boron neutrinos ($E \sim 5-10~\text{MeV}$, $\ell_{\text{osc}} \approx
150-300~\text{km}$), the Earth's layers can be divided into cliffs and valleys satisfying
\eqref{piecewise_continuity}, as well as into a number of layers whose widths are of the order of
the oscillation length, while the density variations are small. One can show that the contributions
of the latter segments can be easily considered in a manner very similar to those of the cliffs,
hence we do not consider them until Sec.~\ref{sec:CrustEffect} where they will be paid special
attention. In the present section, it is enough to note that the valley-cliff classification
depends on the neutrino energy.  It is also worth saying here that during the night, the neutrino
ray traverses different paths through the Earth, thus, the lengths $\epsilon_j$ and $L_j$ vary.
However, assumption \eqref{piecewise_continuity} holds for the most part of the night.

The total flavor evolution operator for our piecewise continuous density profile equals
the matrix product of the evolution operators for all cliffs and valleys. Within each of these
segments, the two small parameters arise: the first of them,
\begin{equation}\label{eq:eta_constraints}
   \eta = \frac{2EV(x)}{\sDelta  m^2} \lesssim
          \begin{cases}
             1.0 \times 10^{-2}, &\qquad E = 0.862~\text{MeV},\\
             1.2 \times 10^{-1}, &\qquad E \sim 10~\text{MeV},
          \end{cases}
\end{equation}
due to the relatively small density of the Earth \cite{IoannisianSmirnov04,IoannisianSmirnov05},
while the second parameter due to the piecewise continuous structure of the density profile,
\begin{equation}\label{delta}
    \delta = \left\{\begin{array}{ll}
                      \ell_{\text{osc}} / L_j & \text{for $j$th valley,}\\
                      \epsilon_j / \ell_{\text{osc}} & \text{for $j$th cliff.}
                    \end{array}
             \right.
\end{equation}
The smallness of parameter $\delta$ will be discussed in Sec.~\ref{sec:Discussion}.

 The evolution matrix for each segment, as well as the total evolution matrix, can be subjected
to the following unitary transformations \cite{Olivo08}:
\begin{eqnarray}\label{eq2}
 R(x,x_{0}) &=& Z^{+}(x)Y(\psi(x))R_{0}(x,x_{0})Y^{-1}(\psi(x_{0}))Z^{-}(x_{0}),\\
 Z^{\pm}(x)&=&\frac{1}{\sqrt{2}}\left\{\sqrt{1-\frac{a(x)}{\omega(x)}}\pm
 \ii\sigma_{2}\sqrt{1+\frac{a(x)}{\omega(x)}}\,\right\},\\
 Y(\psi(x))&=& \cos\psi(x) + \ii\sigma_{3}\sin\psi(x) = \exp\{\ii\sigma_3 \psi(x)\},
\end{eqnarray}
where $\omega(x) = \sqrt{a^{2}(x)+b^{2}}$ 
and $\psi(x) = \lambda \int \omega(x)\diff{x}$ is the corresponding phase incursion. For the
calculations which follow, it is also useful to introduce the effective mixing angle in the medium
$\theta(x) \in [0,\pi/2]$ \ \cite{MikheevSmirnov}, which is defined by the expressions
\begin{equation}\label{theta_x}
  \cos2\theta(x) = -\frac{a(x)}{\omega(x)}, \qquad \sin2\theta(x) = \frac{b}{\omega(x)}.
\end{equation}
In terms of this angle,
\begin{equation}
    Z^\pm(x) = \exp\{\pm\ii\sigma_2 \theta(x)\}.
\end{equation}

The transformation with matrices $Z^\pm(x)$ locally diagonalizes the Hamiltonian $H(x)$ in the
point $x$. It thus makes the complete diagonalization in the homogeneous case $N_e(x) = \const$
\cite{MikheevSmirnov}. The transformation with the operator $Y(\psi(x))$ isolates the effect of the
medium inhomogeneity, the transformed evolution matrix $R_{0}(x,x_{0})$ satisfying the equation
\begin{equation}\label{eq4}
 \frac{\pd R_{0}(x,x_{0})}{\pd x} = -\ii \dot\theta(x) \sigma_2 e^{2\ii\sigma_3 \psi(x)}R_{0}(x,x_{0}),
\end{equation}
where the dot over $\theta$ denotes the gradient
\begin{equation}
    \dot\theta(x) \equiv \pd_x \theta(x) = \frac{b \, \pd_x{a}(x)}{2\omega^2(x)}.
\end{equation}
Due to the fact that the neutrino detector is homogeneous ($\dot\theta(x) = 0$), Eq.~\eqref{eq4}
has a well-defined and physically relevant $x \to +\infty$ limit for any fixed $x_0$. Moreover,
asymptotically convergent behavior of such systems of differential equations is stated by the
Levinson's theorem \cite{DiffEquations}.

In the homogeneous case, the equation above is trivial, $R_0(x,x_0) \equiv \idMatrix$. However, in
the valley $[x_j^+, x_{j+1}^-]$, the slow change of the density $N_e(x)$ enables us to use the
so-called adiabatic approximation leading to the same result \cite{MikheevSmirnov}
\begin{equation}\label{eq:R0_adiab}
  R_0(x_{j+1}^-,x_{j}^+) = \idMatrix + \bigO(\sDelta\eta \ \delta_{\text{valley}}),
\end{equation}
where the remainder term is a (generally speaking, non-diagonal) matrix and $\sDelta\eta$ is the
variation of the density parameter $\eta$ over the valley. On the other hand, if the Wolfenstein
potential undergoes a considerable change within the narrow cliff $[x_{j}^-, x_{j}^+]$ with the
phase incursion $\sDelta\psi \ll 2\pi$, then we get
\begin{equation}
R_{0}(x_j^+,x_j^-) = \exp\left\{-\ii \sigma_2 \sDelta\theta_j e^{2\ii\sigma_3\psi(x_j^-)}\right\} +
\bigO(\eta\delta_{\text{cliff}}),
\end{equation}
where $\sDelta\theta_j \equiv \theta(x_j^+) - \theta(x_j^-) = \bigO(\eta)$ is the jump of the
effective mixing angle on the $j$th cliff. Moreover, one can show that within the more accurate
$O(\eta\delta)$ approximation, the above expressions take the form (see the Appendix)
\begin{eqnarray}\label{eq6}
   R_0(x_{j+1}^-, x_j^+) &=& \exp\left\{- \frac{\ii\sigma_1}{2\lambda} \left[e^{2\ii\sigma_3\psi(x_{j+1}^-)}\dot\theta(x_{j+1}^-) -
                                                                    e^{2\ii\sigma_3\psi(x_{j}^+)}\dot\theta(x_j^+)
                                                                   \right]\right\} + \bigO(\sDelta\eta \ \delta^2) \quad
                                                                   \text{(valley)},
   \\
   R_0(x_j^+, x_j^-) &=& \exp\{ (-\ii\sigma_2\sDelta\theta_j + \ii\sigma_1\mu_j) e^{2\ii\sigma_3\psi(x_j^-)}\} + \bigO(\eta\delta^2)
   \qquad\qquad\qquad\qquad\qquad
   \text{(cliff)} \label{eq7},
\end{eqnarray}
where
\begin{equation}
    \mu_j = 2\lambda\int\limits_{x_j^-}^{x_j^+} (y-x_j^-) \dot\theta(y)\diff{y} = \bigO(\eta\delta).
\end{equation}
\vspace{0.5em}%
Now let us write the evolution operator for the whole neutrino path. The neutrinos observed during
the day are created in the point $x_0$ inside the solar core, then travel to the Earth, enter the
detector in the point $x_1$ and are finally absorbed in the point $x^*$ inside it. In the
nighttime, however, after reaching the Earth in the point $x_1$, the neutrinos pass through a number of
Earth's layers (valleys) discussed above, and only after that they enter the detector in the point
$x_n$ and are absorbed in $x^*$. Crossing the Sun-to-vacuum interface, as well as traveling inside
the Sun, does not involve steep electron density changes, moreover, the neutrino
propagation is highly adiabatic there (see Sec.~\ref{sec:Discussion}), thus we can treat the whole
segment $[x_0,x_1]$ as a single valley and use the adiabatic approximation \eqref{eq:R0_adiab}. As
it was mentioned earlier, the flavor evolution operator for the whole neutrino path is a matrix
product of the evolution operators for each segment (each valley and cliff). By the substitution of
the approximate solutions \eqref{eq6} and \eqref{eq7} into representation \eqref{eq2}, after some
transformations we find the total evolution operator in the form (compare with
Eq.~\eqref{R_valley_cliff})
\begin{eqnarray}\label{eq9}
R(x^*,x_{0}) &=& R_{\text{det}}(x^*,x_{n}^{+})
                 e^{\ii\sigma_2\theta_n^-} e^{\ii\sigma_1(\mu_n - \dot\theta_n^- / 2\lambda)}
                 e^{\ii\sigma_3\sDelta\psi_n}
                 e^{-\ii\sigma_2\sDelta\theta_{n-1}}e^{\ii\sigma_1\bar\mu_{n-1}}
                 e^{\ii\sigma_3\sDelta\psi_{n-1}}
                 e^{-\ii\sigma_2\sDelta\theta_{n-2}}e^{\ii\sigma_1\bar\mu_{n-2}}
                 e^{\ii\sigma_3\sDelta\psi_{n-2}}
                 \nonumber\\
                 &\times&
                 \ldots\times e^{\ii\sigma_3\sDelta\psi_2}
                 e^{-\ii\sigma_2\sDelta\theta_1}e^{\ii\sigma_1\bar\mu_{1}}
                 e^{\ii\sigma_3\sDelta\psi_1}
                 e^{-\ii\sigma_2\theta_{\text{Sun}}} +
                 \bigO(n \eta\delta^2).
\end{eqnarray}
Here, the subscript `Sun' refers to the point $x_0$ inside the solar core, where the neutrino is
created, and the evolution operator inside the neutrino detector is denoted $R_{\text{det}}$. The
factor $n$ in the remainder term indicates that it contains the sum over all cliffs and valleys.
Moreover, we use the following notation:
\begin{gather}
   \theta_j^- \equiv \theta(x_j^-), \quad \sDelta\theta_j \equiv \theta(x_j^+) - \theta(x_j^-), \qquad j = \overline{1,n-1},\\
   \dot\theta_j^- \equiv \dot\theta(x_j^-), \quad \sDelta\dot\theta_j \equiv \dot\theta(x_j^+) - \dot\theta(x_j^-),
   \qquad j = \overline{1,n-1},\\
   \bar\mu_j \equiv \mu_j + \frac{\sDelta\dot\theta_j}{2\lambda} = \int\limits_{x_j^-}^{x_j^+}
   \left(2\lambda(x-x_j^-)\dot\theta(x) + \frac{\ddot\theta(x)}{2\lambda}\right)\diff{x}, \qquad j = \overline{1,n-1}, \label{bar_mu_j_def}\\
   \sDelta\psi_j \equiv \psi(x_j^-) - \psi(x_{j-1}^-) = \lambda \int\limits_{x_{j-1}^-}^{x_j^-}
   \omega(x)\diff{x}, \qquad j = \overline{1,n}.
\end{gather}
It is also convenient to append definition \eqref{bar_mu_j_def} with
\begin{equation}
    \bar\mu_n \equiv \mu_n - \dot\theta_n^- / 2\lambda.
\end{equation}
If the boundary between the Earth's crust and the detector is abrupt, $x_n^+ - x_n^- \ll
\ell_{\text{osc}}$, then $\mu_n$ vanishes. Quite analogously, $\mu_1$ vanishes for the abrupt
vacuum-to-Earth boundary.

\vspace{0.5em}%
By projecting the neutrino state onto the flavor eigenstates, we arrive at the observation
probabilities for the electron/muon neutrino
\begin{equation}\label{eq100}
  P_{e,\mu} \equiv \left\{
    \begin{array}{r}
           P(\nu_e\to\nu_e)\\
           P(\nu_e\to\nu_\mu)
    \end{array}
  \right\} =
\frac{1}{2} \pm \frac{1}{4}\tr\left\{ R(x^*,x_{0})\sigma_{3}R\hc(x^*,x_{0})\sigma_{3}\right\} =
\frac{1 \pm T}{2},
\end{equation}
where, for nighttime neutrinos,
\begin{eqnarray}
T = T_{\text{night}} &=& \frac12 \tr\bigl\{\sigma_3
                                 R_{\text{det}}\hc(x^*,x_n^+) \sigma_3 R_{\text{det}}(x^*,x_n^+)
             e^{\ii \sigma_2\theta_n^-} e^{\ii\sigma_1\bar\mu_n}
             e^{\ii \sigma_3\sDelta\psi_n} e^{-\ii \sigma_2 \sDelta\theta_{n-1}}e^{\ii\sigma_1\bar\mu_{n-1}}
             e^{\ii \sigma_3\sDelta\psi_{n-1}}
  \nonumber\\
  &\times&
             e^{-\ii \sigma_2 \sDelta\theta_{n-2}}e^{\ii\sigma_1\bar\mu_{n-2}}
             e^{\ii \sigma_3\sDelta\psi_{n-2}}
             \ldots e^{-\ii \sigma_2 \sDelta\theta_{1}}e^{\ii\sigma_1\bar\mu_1}
             e^{\ii\sigma_3\sDelta\psi_1}
             e^{-2\ii\sigma_2\theta_{\text{Sun}}}
  \nonumber\\
  &\times&
             e^{-\ii \sigma_3\sDelta\psi_1}
             e^{\ii\sigma_1\bar\mu_1}e^{-\ii \sigma_2 \sDelta\theta_{1}}
             \ldots
             e^{-\ii \sigma_3\sDelta\psi_{n-1}}e^{\ii\sigma_1\bar\mu_{n-1}}
             e^{-\ii \sigma_2 \sDelta\theta_{n-1}}
             e^{\ii\sigma_1\bar\mu_n}
             e^{-\ii \sigma_3\sDelta\psi_n}e^{\ii \sigma_2\theta_n^-}
             \bigr\}, \label{eq101}
\end{eqnarray}
while for daytime neutrinos we have
\begin{equation}\label{eq101b}
T = T_{\text{day}} = \frac12 \tr\bigl\{\sigma_3
             R_{\text{det}}\hc(x^*,x_n^+) \sigma_3 R_{\text{det}}(x^*,x_n^+)
             e^{\ii\sigma_2\theta_0}
             e^{\ii \sigma_3\sDelta\psi_1}
             e^{-2\ii\sigma_2\theta_{\text{Sun}}}
             e^{-\ii \sigma_3\sDelta\psi_1}
             e^{\ii\sigma_2\theta_0}\bigr\}.
\end{equation}
The mixing angle immediately before the detector obviously takes the vacuum value $\theta_0$ in
this case.

Due to the homogeneity of the detector substance and its smallness compared with the oscillation
length, we easily find
\begin{eqnarray}\label{R_det}
  R_{\text{det}}(x^*,x_n^+) = Z_{\text{det}}^+ e^{\ii\sigma_3\sDelta\psi_{\text{det}}}
  Z_{\text{det}}^- &=& e^{\ii\sigma_2\theta_{\text{det}}} (\idMatrix + \ii\sigma_3\sDelta\psi_{\text{det}})
  e^{-\ii\sigma_2\theta_{\text{det}}} + \bigO(\delta_{\text{det}}^2),
  \\
  \sigma_3 R_{\text{det}}\hc(x^*,x_n^+) \sigma_3 R_{\text{det}}(x^*,x_n^+) &=& \idMatrix - 2\ii\sigma_1 \sDelta\psi_{\text{det}}
  \sin2\theta_{\text{det}} + \bigO(\delta_{\text{det}}^2), \label{R_det_squared}
\end{eqnarray}
where the small parameter $\delta_{\text{det}}$ is the ratio of the detector width $L_{\text{det}}$
to the oscillation length $\ell_{\text{osc}}$. The quadratic remainder terms can obviously be
neglected.

\section{Day-night asymmetry}\label{sec:DNA}
\subsection{Finding the probabilities}

In order to evaluate the probabilities obtained above, let us first make the averaging over the
phase $\sDelta\psi_1$, which corresponds to the neutrino path between the creation point $x_0$ and
the Earth. The region of the neutrino creation is extremely large compared with the oscillation
length, thus, after this averaging, $\avg{\cos 2\sDelta\psi_1} = \avg{\sin{ 2\sDelta\psi_1}} = 0$
with a high accuracy. Using this fact together with Eq.~\eqref{R_det_squared}, after averaging
\eqref{eq101b} we find
\begin{eqnarray}\label{eq:T_avg_deltaPsi1}
  \avg{e^{\ii\sigma_3\sDelta\psi_1}  e^{-2\ii\sigma_2\theta_{\text{Sun}}}
  e^{-\ii\sigma_3\sDelta\psi_1}} &=& \cos2\theta_{\text{Sun}},\\
  \avg{T_{\text{day}}} &=& \cos2\theta_0 \cos2\theta_{\text{Sun}}. \label{eq:Tday_avg}
\end{eqnarray}
The latter expression constitutes the famous result of Mikheev and Smirnov \cite{MikheevSmirnov}.
On the other hand, after averaging over $\sDelta\psi_1$
for the nighttime neutrinos, we obtain
\begin{eqnarray}
  T_{\text{night}} &\rightarrow& \frac12\cos2\theta_{\text{Sun}} \tr\bigl\{
             (e^{2\ii\sigma_2\theta_n^-} - 2\ii\sigma_1 \sDelta\psi_{\text{det}}\sin2\theta_{\text{det}})
             e^{\ii\sigma_1\bar\mu_n}e^{\ii \sigma_3\sDelta\psi_n}
             e^{-\ii \sigma_2 \sDelta\theta_{n-1}}e^{\ii\sigma_1\bar\mu_{n-1}}
             e^{\ii \sigma_3\sDelta\psi_{n-1}}
             \ldots \bigr.\nonumber\\
   &\times&  e^{\ii \sigma_3\sDelta\psi_2}
             e^{-2\ii \sigma_2\sDelta\theta_1}e^{2\ii\sigma_1\bar\mu_1}
             e^{-\ii \sigma_3\sDelta\psi_2}\ldots
             e^{-\ii \sigma_3\sDelta\psi_{n-1}}
             e^{-\ii \sigma_2 \sDelta\theta_{n-1}} e^{\ii\sigma_1\bar\mu_{n-1}}
             e^{-\ii \sigma_3\sDelta\psi_n} e^{\ii\sigma_1\bar\mu_n}
             \bigr\}, \label{eq:T_night_avg}
\end{eqnarray}
where we have made use of the fact that matrices $e^{\ii\sigma_1\bar\mu_1}$ and
$e^{-\ii\sigma_2\sDelta\theta_1}$ commute up to a negligible term of the order
$\bigO(\sDelta\theta_1\bar\mu_1)$.

For the calculations which follow, we will use the smallness of the jumps
$\sDelta\theta_1,\ldots,\sDelta\theta_{n-1}=\bigO(\eta)$ and the parameters
$\bar\mu_1,\ldots,\bar\mu_n=\bigO(\eta\delta)$. Within the linear approximation in the Earth's
density parameter $\eta$, this leads to
\begin{equation}\label{eq:T_series}
   T_{\text{night}}(\sDelta\theta_1,\ldots,\sDelta\theta_{n-1}; \bar\mu_1,\ldots,\bar\mu_n) =
   \cos2\theta_n^-\cos2\theta_{\text{Sun}}
   + \sum\limits_{j=1}^{n-1}
   \sDelta\theta_j\cdot
   \left.\frac{\pd T_{\text{night}}}{\pd(\sDelta\theta_j)}\right|_{\sDelta\theta,\bar\mu=0}
   + \sum\limits_{j=1}^{n} \bar\mu_j \cdot
   \left.\frac{\pd T_{\text{night}}}{\pd\bar\mu_j}\right|_{\sDelta\theta,\bar\mu=0}.
\end{equation}
Partial derivatives with respect to the small parameters are
\begin{eqnarray}
    \left.\frac{\pd T_{\text{night}}}{\pd(\sDelta\theta_j)}\right|_{\sDelta\theta,\bar\mu = 0} &=&
    -\ii\cos2\theta_{\text{Sun}} \tr\{(\sigma_2 e^{2\ii\sigma_2\theta_n^-} + 2\sigma_3 \sDelta\psi_{\text{det}}
  \sin2\theta_{\text{det}}) e^{-2\ii\sigma_3\sDelta\psi_{n,j}}\}
  \nonumber\\
  &=&
   2\cos2\theta_{\text{Sun}} \left\{ \sin2\theta_n^- \cos2\sDelta\psi_{n,j}
   - 2\sDelta\psi_{\text{det}} \sin2\theta_{\text{det}} \sin2\sDelta\psi_{n,j}\right\},\label{eq:T1}
   \\
   \left.\frac{\pd T_{\text{night}}}{\pd\bar\mu_j}\right|_{\sDelta\theta,\bar\mu = 0} &=&
   \cos2\theta_{\text{Sun}} \tr\{(\ii e^{2\ii\sigma_2\theta_n^-}\sigma_1 + 2 \sDelta\psi_{\text{det}}
   \sin2\theta_{\text{det}}) e^{-2\ii\sigma_3\sDelta\psi_{n,j}}\}
   \nonumber\\
   &=&
   2\cos2\theta_{\text{Sun}} \left\{ \sin2\theta_n^- \sin2\sDelta\psi_{n,j}
   + 2\sDelta\psi_{\text{det}} \sin2\theta_{\text{det}} \cos2\sDelta\psi_{n,j}\right\}.
   \label{eq:T11}
\end{eqnarray}
In the above expressions,
\begin{equation}\label{eq:T2}
   \sDelta\psi_{n,j} \equiv \psi(x_{n}^-)-\psi(x_j^-)
                     =      \lambda \int\limits_{x_j^-}^{x_n^-} \omega(x) \mathrm{d}x
                     =      \lambda L_{n,j} (1+ \bigO(\eta)),
\end{equation}
where $L_{n,j} \equiv x_n^- - x_j^-$ is the distance between the boundary of the $j$th crossed
Earth's shell and the detector, measured along the neutrino ray. Finally, by substituting the
derivatives \eqref{eq:T1} and \eqref{eq:T11} into Eq.~\eqref{eq:T_series} and using the fact that
$\sin2\sDelta\psi_{n,n} = 0$, we arrive at the final result
\begin{eqnarray}
    T_{\text{night}} &=& \cos2\theta_{\text{Sun}}
        \Bigl\{
                \cos2\theta_n^- +
                2\sin2\theta_n^- \sum_{j=1}^{n-1} (\sDelta\theta_j \cos2\sDelta\psi_{n,j} +
                                                   \bar\mu_j \sin2\sDelta\psi_{n,j})
    \nonumber\\
        &&\qquad\qquad - 4\sDelta\psi_{\text{det}} \sin2\theta_{\text{det}} \sum_{j=1}^{n-1}
        \sDelta\theta_j \sin2\sDelta\psi_{n,j}
        \Bigr\},
    \label{eq:T_approx}
\end{eqnarray}
which is valid up to the terms of the order $\bigO(\eta\delta^2)$ and quadratic terms
$\bigO(\eta^2)$. The term including the product of the detector width $\sDelta\psi_{\text{det}}$
and the oscillating factor $\bar\mu_j\cos2\sDelta\psi_{n,j}$ is of the order
$\bigO(\eta\delta\delta_{\text{det}})$ and is thus omitted.

The above expression provides a generalization of the main result of paper \cite{Wei} for the case
of nonzero-thickness cliffs and detector. It should be stressed, however, that
Eq.~\eqref{eq:T_approx} gives poor information on the effects to be measured. Indeed, the neutrino
experiments last for years, and thus, Eq.~\eqref{eq:T_approx} may only acquire a predictive power
after some kind of averaging. The averaging procedure should take into account the axial rotation
of the Earth (involving the integration over the nights), as well as its orbital motion around the
Sun.

\subsection{Averaging the probabilities}\label{sec:Averaging}

The averaging procedure can be performed analytically, if the oscillation phase incursions
$\sDelta\psi_{n,j}$ vary by much more than $2\pi$ during the night. Namely, in this case, one can
employ the stationary phase technique (see, e.g., \cite{StationaryPhase}). In the case of the
beryllium neutrinos ($E = 0.862$~MeV) traveling through the Earth, the oscillation length is about
30km, while the depths  of the valleys $L_{n,j}$ vary by many hundreds of kilometers, so
the oscillation phase variations are indeed large enough to use the stationary phase approximation.
To some extent, the same holds for boron neutrinos. However, for both neutrino types,
there are layers, to which the stationary phase approximation may not apply. These are the Earth's
crust immediately under the detector and (for boron neutrinos) the upper mantle. These layers are
discussed in detail in section~\ref{sec:CrustEffect} and Appendix~\ref{app:crust} and do not
interfere with the picture described in this paragraph.

\begin{figure}
  \includegraphics[width=12cm]{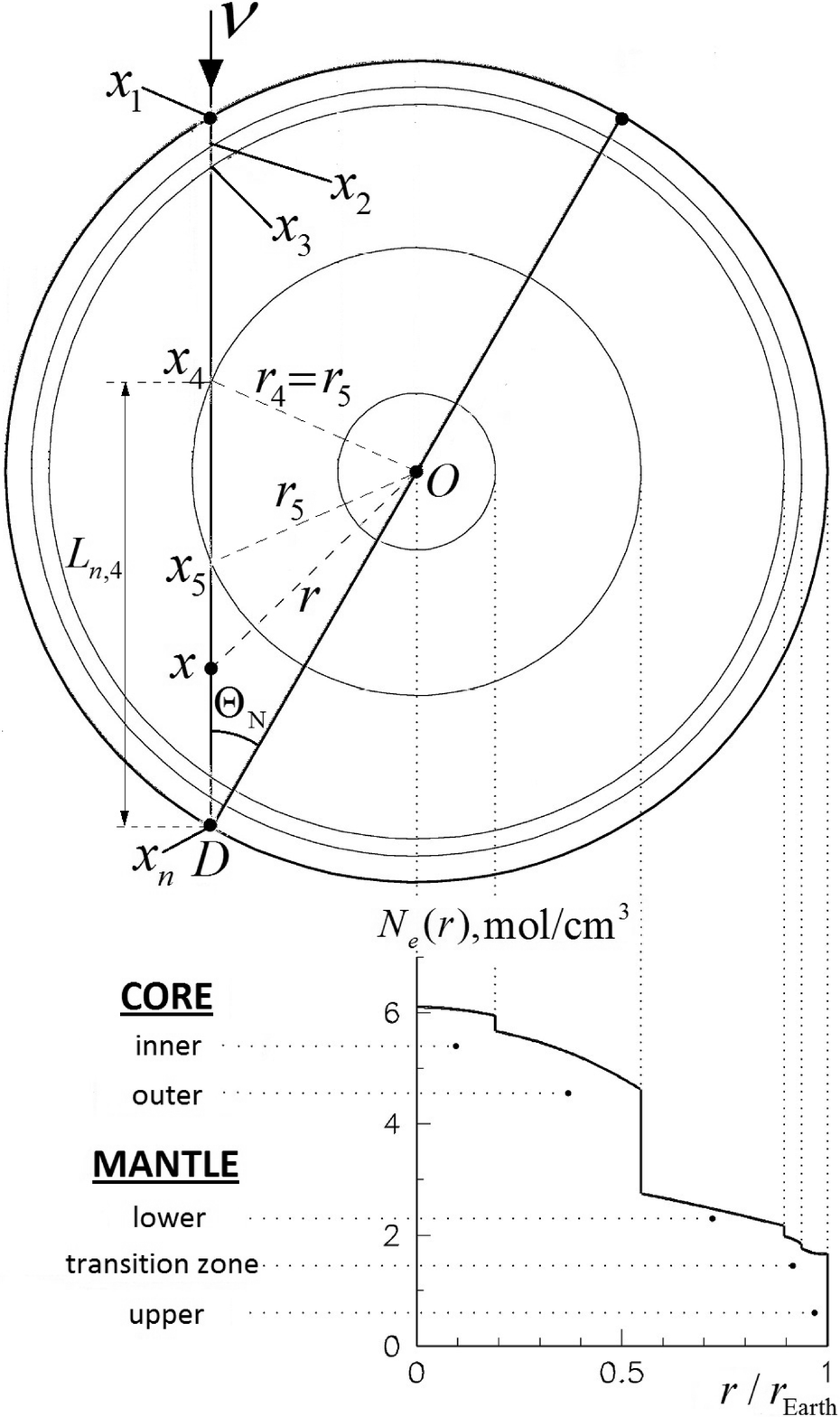}
  \caption{Radial distribution of the electron density $N_e(r)$ inside the Earth and the neutrino
  path through it. The figure demonstrates the cross section of the Earth which contains the nadir $DO$,
  the center of the Earth $O$, and the neutrino ray}
  \label{fig:EarthDensity}
\end{figure}

Let us consider a neutrino traveling through the Earth, which, according to the PREM model
\cite{Anderson}, consists of a number of concentric spherical shells. The boundary between the
valleys $x_j$ corresponds to the point where the neutrino crosses one of the interfaces between the
Earth's shells; let $r_j$ be the radius of this interface (see Fig.~\ref{fig:EarthDensity}).
Further, the distances $L_{n,j}^\pm$ between the detector and the points where the neutrino
enters/leaves the interface with radius $r_j$ are functions of the `nadir angle' $\Theta_{\text{N}}
\in [0,\pi]$ defined as the angle between the direction to the Sun and the nadir in the point of
the detector. In terms of the solar elevation angle $\Theta_{\text{s}}$ \cite{Astronomy}, the nadir
angle is $\Theta_{\text{N}} = \Theta_{\text{s}} + \pi/2$. The nadir angle, in turn, is a function
of the Earth's axial rotation angle $\tau \in [0,2\pi)$ (`time of day') and the orbital motion
angle $\varsigma \in [0,2\pi)$ (`season').  The dependence of the distances on the nadir angle is
easily found to be
\begin{eqnarray}\label{eq:Lnj_ThetaZ}
    L_{n,j} &=& L_{n,j}^{\pm}(\Theta_{\text{N}}) = r_n\cos\Theta_{\text{N}} \pm \sqrt{r_j^2 - r_n^2
    \sin^2\Theta_{\text{N}}}, \\
    \Theta_{\text{N}} &\le& \arcsin r_j / r_n,  \label{eq:Lnj_ThetaZ_condition}
\end{eqnarray}
where the upper/lower signs in \eqref{eq:Lnj_ThetaZ} correspond to the neutrino entering/leaving
the interface $r_j$ (see Fig.~\ref{fig:EarthDensity}). The inequality
\eqref{eq:Lnj_ThetaZ_condition} ensures that the intersection of the neutrino ray with this
interface exists.

In order to find the night average of the electron/muon neutrino observation probabilities, let us
note some properties of expressions \eqref{eq:T_approx} and \eqref{eq:Lnj_ThetaZ}. First, the
number of interfaces crossed by the neutrino is defined via the inequality
\eqref{eq:Lnj_ThetaZ_condition}, so the number of the valleys and, thus, the number of terms
entering the sums in \eqref{eq:T_approx} are changing during the night.
Therefore, the night average of Eq.~\eqref{eq:T_approx} contains the sum over all interfaces $j$,
with the average of the $j$th oscillating term involving $\sDelta\psi_{n,j}$ defined as follows:
\begin{equation}\label{eq:night_averaging}
    \avg{F(\Theta_{\text{N}})e^{2\ii\sDelta\psi_{n,j}}}_{\text{night}} =
    \int\limits_{\Theta_{\text{N}}(\tau) \le \arcsin r_j/r_n} \!\!\!\!\!\!\!\!\!\!\!\!
    F(\Theta_{\text{N}}(\tau))e^{2\ii\sDelta\psi_{n,j}(\Theta_{\text{N}}(\tau))} \; \frac{\diff\tau}{\sDelta\tau_{\text{night}}}.
\end{equation}
Here, $F(\Theta_{\text{N}})$ is some slowly changing function of the nadir angle and
$\sDelta\tau_{\text{night}}$ is the total duration of the night in terms of the Earth's axial
rotation angle $\tau$, namely, the length of the segment where $\Theta_{\text{N}}(\tau) < \pi/2$
(the Sun is below the horizon).

Second, the duration of the night is, in turn, a function of the season $\varsigma$. On the
equinox, e.g., $\sDelta\tau_{\text{night}} = \pi$, while on the winter solstice,
$\sDelta\tau_{\text{night}} \to \max$. However, the summer nights are just as long as the opposite
winter days, so that
\begin{equation}
    \sDelta\tau_{\text{night}}(\varsigma + \pi) = 2\pi - \sDelta\tau_{\text{night}}(\varsigma),
\end{equation}
and the total duration of the nights over all the year is exactly half the year. Therefore, the
averaging over the year of $N_\varsigma$ days should involve the division by the total duration of
the nights, i.e., $\pi N_\varsigma$. For $N_\varsigma \gg 1$, the summation over the nights can be
replaced by the integration,
\begin{equation}
    \avg{\ldots}_{\text{night,year}} = \frac{1}{N_\varsigma\pi} \sum\limits_{\varsigma = \varsigma_k}
    \int\diff\tau (\ldots)
                                     \approx \frac{1}{2\pi^2}\int\limits_0^{2\pi} \diff\varsigma
                                     \int \diff\tau (\ldots), \qquad \varsigma_k = \frac{2\pi k}{N_\varsigma},
                                     k=1,\ldots,N_\varsigma,
\end{equation}
and the averaging formula for the terms containing the phase incursion $\sDelta\psi_{n,j}$ takes
the form
\begin{equation}
    \avg{F(\Theta_{\text{N}})e^{2\ii\sDelta\psi_{n,j}}}_{\text{night,year}} =
    \frac{1}{2\pi^2}\int\limits_{0}^{2\pi} \diff\varsigma
    \int\limits_{\Theta_{\text{N}}(\tau,\varsigma) \le \arcsin r_j/r_n} \!\!\!\!\!\!\!\!\!\!\!\!
    \diff\tau \;
    F(\Theta_{\text{N}}(\tau,\varsigma))\,
    e^{2\ii\sDelta\psi_{n,j}(\Theta_{\text{N}}(\tau,\varsigma))}.
\end{equation}

Now we are able to apply the stationary phase technique to such an integral containing the rapidly
oscillating exponential. Indeed, let us use the expression, which is valid for smooth functions
$f(x)$ and $S(x)$ defined on segment $[a,b]$ containing a single non-degenerate stationary point
$x_0 \in (a,b)$ such that $S'(x_0) = 0$, $S''(x_0) \ne 0$ \cite{StationaryPhase}
\begin{equation}
    \int\limits_a^b f(x) e^{\ii \lambda S(x)} \diff{x} = \sqrt{\frac{2\pi}{\lambda |S''(x_0)|}}
    f(x_0)
    e^{\ii\lambda S(x_0) + \ii(\pi/4)\sgn S''(x_0)}
    + \left.\frac{f(y)e^{\ii\lambda S(y)}}{\ii\lambda S'(y)}\right|_a^b + \bigO(\lambda^{-3/2}),
    \quad \lambda \to +\infty.
\end{equation}
The two leading terms come from the stationary point and the boundary, respectively. However, in
the application to the integral \eqref{eq:night_averaging}, the boundary term vanishes. Indeed, the
boundary of the integration domain corresponds to the neutrino ray being tangent to the interface
with radius $r_j$, hence, $\pd_\tau \sDelta\psi_{n,j} \propto \pd_\tau
L_{n,j}(\Theta_{\text{N}}(\tau)) \to \infty$, and the boundary term is absent. On the other hand,
the stationary point is obviously achieved at midnight, when the nadir angle $\Theta_{\text{N}} \to
\min$ (the Sun is in its lowest position), so the integration over the night yields
\begin{equation}\label{eq:night_intergration_general}
    \int\limits_{\Theta_{\text{N}}(\tau) \le \arcsin r_j/r_n} \!\!\!\!\!\!\!\!\!\!\!\!
    F(\Theta_{\text{N}}(\tau))\,
    e^{2\ii\sDelta\psi_{n,j}(\Theta_{\text{N}}(\tau))} \; \frac{\diff\tau}{\pi} =
    \left.\sqrt{\frac{1}{\pi\lambda |\pd_\tau^2 L_{n,j}|}}
    F(\Theta_{\text{N}})
    e^{2\ii\sDelta\psi_{n,j}\mp \ii\pi/4}\right|_{\text{midnight}}
    + \bigO((\lambda L_{n,j})^{-\frac32}),
\end{equation}
where the two possible signs before $\ii\pi/4$ correspond to $L_{n,j} = L_{n,j}^\pm$ (see
Eq.~\eqref{eq:Lnj_ThetaZ}). The principal point here is that the second derivative $\pd_\tau^2
L_{n,j}$ at midnight is suppressed for the inner Earth's shells,
\begin{eqnarray}
  \left[\pd_\tau^2 L_{n,j}(\Theta_{\text{N}}(\tau))\right]_{\text{midnight}} &=&
  \left[\pd_\tau^2(\cos\Theta_{\text{N}}) \frac{\diff
  L_{n,j}(\Theta_{\text{N}})}{\diff(\cos\Theta_{\text{N}})}\right]_{\text{midnight}},\\
  \left[\frac{\diff L_{n,j}(\Theta_{\text{N}})}{\diff(\cos\Theta_{\text{N}})}\right]_{\text{midnight}}
   &=& \pm \left[\frac{L_{n,j}(\Theta_{\text{N}})}{\sqrt{r_j^2/r_n^2 -
   \sin^2\Theta_{\text{N}}}}\right]_{\text{midnight}}.
\end{eqnarray}
Now let us use the expression of the nadir angle $\Theta_{\text{N}}$ via the Earth's axial tilt
$\varepsilon = 23.5\degree$, the latitude of the detector $\chi \in [-\pi/2,\pi/2]$, and the season
$\varsigma \in [0,2\pi]$ \cite{Astronomy},
\begin{equation}
    \cos\Theta_{\text{N}}(\tau,\varsigma) = \cos\chi \sin\varsigma \sin\tau + \cos\varepsilon
    \cos\chi \cos\varsigma \cos\tau + \sin\varepsilon \sin\chi \cos\varsigma,
\end{equation}
where $\varsigma = 0$ corresponds to the winter solstice in the northern hemisphere. The
minimum values of $\Theta_{\text{N}}$ are achieved at midnights corresponding to $\tau =
\tau_{\text{midnight}}$,
\begin{equation}\label{eq:tau_midnight}
    \tan\tau_{\text{midnight}}(\varsigma) = \frac{\tan\varsigma}{\cos\varepsilon}, \quad
    \cos\tau_{\text{midnight}}(\varsigma) = \sgn(\cos\varsigma)\frac{\cos\varepsilon}{\sqrt{\cos^2\varepsilon + \tan^2\varsigma}}.
\end{equation}
Using these expressions, we find the derivative of $\cos\Theta_{\text{N}}$ at midnight,
\begin{equation}
    \left[\pd_\tau^2(\cos\Theta_{\text{N}})\right]_{\text{midnight}} = -
    \frac{\cos\chi}{|\cos\varsigma|} \,\frac{\cos^2\varepsilon
    +\sin^2\varsigma\sin^2\varepsilon}{\sqrt{\cos^2\varepsilon + \tan^2\varsigma}} \equiv -\mathcal{N}(\varsigma).
\end{equation}
Finally, the integral over the night \eqref{eq:night_intergration_general} takes the form
\begin{equation}\label{eq:after_night_averaging}
    \int\limits_{\Theta_{\text{N}}(\tau) \le \arcsin r_j/r_n} \!\!\!\!\!\!\!\!\!\!\!\!
    F(\Theta_{\text{N}}(\tau))\,
    e^{2\ii\sDelta\psi_{n,j}(\Theta_{\text{N}}(\tau))} \; \frac{\diff\tau}{\pi} \approx
    \frac{1}{\sqrt{\mathcal{N}(\varsigma)}}
    \left[
    \frac{(r_j^2/r_n^2 - \sin^2\Theta_{\text{N}})^{1/4}}{\sqrt{\pi\lambda
    L_{n,j}(\Theta_{\text{N}})}}
    F(\Theta_{\text{N}})
    e^{2\ii\sDelta\psi_{n,j}\mp \ii\pi/4}
    \right]_{\text{midnight}}.
\end{equation}
On the other hand, the midnight stationary (minimum) values of the nadir angle
$\Theta_{\text{N}}(\tau_{\text{midnight}})$ vary throughout the year (see
Eq.~\eqref{eq:tau_midnight}), being the smallest on the winter solstice (the darkest midnight) and
the largest on the opposite summer solstice (the lightest midnight). Therefore, the right side of
Eq.~\eqref{eq:after_night_averaging} is still containing a rapidly oscillating function of the
season $\varsigma$, and we can perform another isolation of the stationary points, namely, of the
two solstices $\varsigma = 0,\pi$:
\begin{gather}
  \avg{F(\Theta_{\text{N}})e^{2\ii\sDelta\psi_{n,j}}}_{\text{night,year}} =
  \int\limits_0^{2\pi}\frac{\diff\varsigma}{2\pi}
  \int\limits_{\Theta_{\text{N}}(\tau,\varsigma) \le \arcsin r_j/r_n} \!\!\!\!\!\!\!\!\!\!\!\!
    F(\Theta_{\text{N}}(\tau,\varsigma))\,
    e^{2\ii\sDelta\psi_{n,j}(\Theta_{\text{N}}(\tau,\varsigma))} \; \frac{\diff\tau}{\pi}
  \nonumber\\
  =
  \frac{1}{2\pi}\sum\limits_{\varsigma=0,\pi}
  \frac{\vartheta\big(r_j - r_n\sin\Theta_{\text{N}}\big)}{\sqrt{\mathcal{N}(\varsigma)|\pd_\varsigma^2\cos\Theta_{\text{N}}(\tau_{\text{midnight}}(\varsigma))|}}
    \left[
    \frac{\sqrt{r_j^2/r_n^2 - \sin^2\Theta_{\text{N}}}}{\lambda L_{n,j}(\Theta_{\text{N}})}
    F(\Theta_{\text{N}})
    e^{2\ii\sDelta\psi_{n,j} + \ii s'(s-1) \pi/4}
    \right]_{\text{midnight}},
    \label{eq:YearAverage_gen}
    \\
    \pd_\varsigma^2\cos\Theta_{\text{N}}(\tau_{\text{midnight}}(\varsigma)) =
      \begin{cases}
         \sin(\varepsilon - \chi)\tan\varepsilon, & \varsigma = 0 \text{ (winter solstice),}\\
         \sin(\varepsilon + \chi)\tan\varepsilon, & \varsigma = \pi \text{ (summer solstice),}
      \end{cases}\\
    s \equiv \sgn\bigl\{\pd_\varsigma^2\cos\Theta_{\text{N}}(\tau_{\text{midnight}}(\varsigma))\bigr\} =
      \begin{cases}
         -1, & \varsigma = 0,\\
         +1, & \varsigma = \pi,
      \end{cases}\\
    s' = \sgn\bigl\{ L_{n,j}(\Theta_{\text{N}}) - r_n \cos\Theta_{\text{N}} \bigr\} = \pm1 \qquad \text{for } L_{n,j} = L_{n,j}^\pm,\\
    \vartheta(\xi) \equiv       \begin{cases}
                                   1, & \xi \ge 0,\\
                                   0, & \xi < 0.
                                \end{cases}
\end{gather}
In the above expressions, the Heaviside theta function $\vartheta\big(r_j -
r_n\sin\Theta_{\text{N}}\big)$ ensures that the stationary point for the $j$th interface is really
reached, while the additional sign $s'$ is explicitly introduced to avoid $\pm$-expressions
originating from Eq.~\eqref{eq:Lnj_ThetaZ}. The signs specified in Eq.~\eqref{eq:YearAverage_gen}
are valid in the northern non-tropical and non-polar latitudes $\chi \in (\varepsilon,
\pi/2-\varepsilon)$. Indeed, the typical neutrino detectors (SNO, Borexino, Super-Kamiokande) are
situated in temperate latitudes. Here, the midnight solstice nadir angle is
$\Theta_{\text{N}}(\tau_{\text{midnight}}) = \chi + s\varepsilon$ and the prefactor
$\mathcal{N}(\varsigma) = \cos\varepsilon \cos\chi$. With the use of this fact together with the
above general averaging formula, one finally arrives at the year average of the night transition
probability \eqref{eq:T_approx},
\begin{eqnarray}
    \avg{T_{\text{night}}}_{\text{night,year}} &\approx& \cos2\theta_{\text{Sun}} \cos2\theta_n^- + 2\cos2\theta_{\text{Sun}}\sin2\theta_n^-
                                   \sum\limits_{j=1}^{n-1} \sDelta\theta_j \nonumber\\
                 &\times&
                                   \sum\limits_{s=\pm1}
                                   \frac{\vartheta\big(r_j - r_n\sin(\chi+s\varepsilon)\big)}{2\pi\sqrt{\sin\varepsilon \cos\chi \sin(\chi+s\varepsilon)}}
                                   \frac{\sqrt{r_j^2/r_n^2 - \sin^2(\chi+s\varepsilon)}}{\lambda L_{n,j}(\chi+s\varepsilon)}
                                   \cos\{2\sDelta\psi_{n,j} + s'(s-1) \pi/4\},
                  \label{eq:YearAverage}
                                   \\
    s' &\equiv& \sgn\{L_{n,j}(\chi+s\varepsilon) - r_n \cos(\chi+s\varepsilon)\}.
\end{eqnarray}
In the above expression, we have omitted the terms resulting from averaging the $\bigO(\eta\delta)$
terms in Eq.~\eqref{eq:T_approx}, since they are extremely small. The phase incursions
$\sDelta\psi_{n,j}$ should obviously be taken at $\Theta_{\text{N}} = \chi+s\varepsilon$, i.e. at
solstice midnights. For the Borexino detector situated in the Gran Sasso laboratory, with $\theta =
+42.5\degree$, the prefactor in \eqref{eq:YearAverage} which does not depend on $j$ amounts to be
\begin{equation}\label{YearAverage_prefactor}
    \frac{1}{2\pi\sqrt{\sin\varepsilon \cos\chi \sin(\chi + s\varepsilon)}} \approx
    \begin{cases}
       0.51, & \text{ winter solstice ($s=-1$)},\\
       0.31, & \text{ summer solstice ($s=+1$)}.
    \end{cases}
\end{equation}
For the Super-Kamiokande detector, $\theta = +36.2\degree$, and the prefactor equals
$0.60$ and $0.30$ for the two stationary points, respectively.

Let us briefly note that, in the tropical latitudes $|\chi| < \varepsilon$, additional
stationary points appear; in particular, on the Equator $\chi = 0$, they correspond to the
equinoxes. Two additional points meet on the winter solstice, when $\chi \to +\varepsilon$ (on the
Tropic), and one encounters a degenerate stationary point \cite{StationaryPhase}. Although
we have found the analytical expressions for the year averages in the tropical and equatorial zones ($|\chi| \le \varepsilon$),
we do not present them here due to their mathematical complexity and to the fact that the actual
neutrino detectors are situated in the temperate latitudes. We confine ourselves to saying that,
from \eqref{YearAverage_prefactor}, one can infer the amplification of the winter solstice contribution,
as one approaches the Tropic.

One should also mention that, for low-energy neutrinos ($E \lesssim 1~\text{MeV}$), in certain
seasons, the phase incursions $\sDelta\psi_{n,j}$ for a certain layer $j$ may differ by
approximately a multiple of $2\pi$ on the successive nights. As a result, the contributions of
these nights will not cancel each other, quite similarly to those of the nights near the solstices.
This may be called a parametric resonance and, in principle, will lead to local anomalies in the
observed neutrino flux, but, as one may see from Sec.~\ref{sec:Discussion}, the day-night asymmetry
for low-energy neutrinos is quite small and its anomalies are even more challenging to observe.
Thus, we do not pay these additional effects much attention here.

Finally, we are left with the following conclusion. The terms entering Eq.~\eqref{eq:T_approx},
which contain the oscillating functions of the phase incursions $2\sDelta\psi_{n,j}$, are
suppressed as $\bigO\left(\frac{r_j}{r_n\lambda L_{n,j}}\right) =
\bigO\left(\frac{r_j}{r_n}\delta\right)$ after averaging over the year; within the leading
approximation, the resulting averages \eqref{eq:YearAverage} come from the stationary phase points
achieved on the winter and the summer solstices. The suppression becomes stronger for the inner
Earth's shells.

It is spectacular that all the terms of the order $\bigO(\eta\delta)$ in the expression
\eqref{eq:T_approx}, including the one corresponding to the detector, become $\bigO(\eta \delta^2
r_j/r_n)$ after the averaging. The terms of the order $\bigO(\eta)$, which are proportional to
$\sDelta\theta_j$, become $\bigO(\eta \delta r_j/r_n)$, respectively. Finally, we are left with the
average value
\begin{equation}
    \avg{T_{\text{night}}} = \cos2\theta_{\text{Sun}} \cos2\theta_n^- + \bigO\left(n \eta \delta
    \frac{r_j}{r_n}\right).
\end{equation}
By substituting this result together with the daytime average value \eqref{eq:Tday_avg} into
expression \eqref{eq100} for the neutrino observation probabilities, we arrive at the day-night
asymmetry factor
\begin{equation}\label{Adn}
  A_{\text{dn}} \equiv \frac{2(\avg{P_{e,\text{night}}} -
  \avg{P_{e,\text{day}}})}{\avg{P_{e,\text{night}}} + \avg{P_{e,\text{day}}}} =
  -\frac{T_{\text{MSW}}}{1+T_{\text{MSW}}} \cdot \frac{\sin^2 2\theta_0}{\cos2\theta_0}\frac{2 E
  V(x_n^-)}{\sDelta m^2}
  + \bigO\left(n \eta \delta \frac{r_j}{r_n}\right),
\end{equation}
where $T_{\text{MSW}} = \cos2\theta_0 \cos2\theta_{\text{Sun}} =
{a_{\text{vac}}a_{\text{Sun}}}/{\omega_{\text{Sun}}}$ defines the observation probabilities for the
solar neutrinos due to the Mikheev--Smirnov--Wolfenstein effect \cite{MikheevSmirnov} and
$V(x_n^-)$ is the Wolfenstein potential in the Earth under the detector.

\vspace{0.5em}

One should hold in mind that the asymmetry factor defined as \eqref{Adn} may not be
directly measurable in the neutrino experiments, depending on the detection mechanism. Indeed,
definition \eqref{Adn} differs from the one preferred by the experimentalists, the latter being
\begin{equation}\label{Adn_exp}
    A_{\text{dn}}^{\text{(exp)}} = \frac{2(N_{\text{night}} - N_{\text{day}})}{N_{\text{night}} +
    N_{\text{day}}},
\end{equation}
where $N_{\text{day,night}}$ is the number of neutrino events observed during the day/night. These
numbers may not correspond to the electron neutrino fluxes. For example, scattering experiments,
which cannot separate the charged and the neutral current interactions, are unable to directly
measure the electron neutrino flux.
To compare prediction \eqref{Adn} with such experiments, one should reinterpret the observed event
rates $N_{\text{day,night}}$ in terms of the electron neutrino fluxes using some theoretical
assumptions.

Let us provide an example of the connection between the day-night asymmetry factors defined as
\eqref{Adn} and \eqref{Adn_exp}. Namely, in the case of the neutrino-electron scattering
experiment, such as Borexino, the incident neutrinos produce recoil electrons inside the detector,
and the scattering cross sections for such processes are well-known \cite{Nu_e_CrossSection,
Bahcall_CrossSections}, including one-loop corrections \cite{Nu_e_RadCorr}.
The resulting ratio of the total electron/muon neutrino detection cross sections is a function of
the neutrino energy $E$ (as well as of the threshold $T_{\text{min}}$ of the recoil electron
detection). For monochromatic beryllium neutrinos and the actual Borexino's threshold, this ratio
is \cite{Borexino1}
\begin{equation}
    \sigma(\nu_e;E) / \sigma(\nu_\mu;E) \approx 4.5, \qquad E = 0.862~\text{MeV},
\end{equation}
thus, the ratio of the event rates is
\begin{equation}
    \frac{N_{\text{day}}}{N_{\text{night}}} = \frac{\avg{P_{e,\text{day}}} \sigma(\nu_e) + (1-
    \avg{P_{e,\text{day}}}) \sigma(\nu_\mu)}{\avg{P_{e,\text{night}}} \sigma(\nu_e) + (1-
    \avg{P_{e,\text{night}}}) \sigma(\nu_\mu)}.
\end{equation}
The `experimental' day-night asymmetry factor is then easily found to be
\begin{equation}
    A_{\text{dn}}^{\text{(exp)}} 
                                 \approx 
                                         A_{\text{dn}} \; \frac{(\sigma(\nu_e)-\sigma(\nu_\mu))(1+T_{\text{MSW}})}
                                                               {(\sigma(\nu_e)-\sigma(\nu_\mu))(1+T_{\text{MSW}}) + 2\sigma(\nu_\mu)}
\end{equation}
and, for beryllium neutrinos ($T_{\text{MSW}} \approx 0.09$), we obtain
\begin{equation}
    A_{\text{dn}}^{\text{(exp)}} \approx 0.66 A_{\text{dn}}  \qquad (E = 0.862~\text{MeV}).
\end{equation}
For boron and other types of neutrinos, which have continuous energy spectrum, the expression for
the `experimental' day-night asymmetry factor involves the integration over the neutrino energy,
\begin{equation}\label{Adn_exp_continuous}
    A_{\text{dn}}^{\text{(exp)}} = \frac{\int \rho(E)\diff{E}  \times  A_{\text{dn}}(E) \; (\sigma(\nu_e;E)-\sigma(\nu_\mu;E)) (1+T_{\text{MSW}}(E))}
                                        {\int \rho(E)\diff{E}  \times  \{ (\sigma(\nu_e;E)-\sigma(\nu_\mu;E)) (1+T_{\text{MSW}}(E)) + 2 \sigma(\nu_\mu;E)\} }.
\end{equation}
Here, the cross sections $\sigma(\nu_{e,\mu};E)$, the `physical' asymmetry factor $A_{\text{dn}}$, and the Mikheev--Smirnov--Wolfenstein factor $T_{\text{MSW}}$ depend on the neutrino energy, and
$\rho(E)$ is the normalized energy distribution of incident neutrinos. As one can see, the
integration can be easily undertaken numerically using our asymmetry prediction \eqref{Adn} and the
expressions for the effective total cross sections, if the recoil electron detection threshold
$T_{\text{min}}$ is known (see, e.g., \cite{Bahcall_CrossSections}).

For those experiments which directly observe the charged-current electron neutrino events, one formally sets $\sigma(\nu_\mu;E) \to 0$. Then one has
$A_{\text{dn}}^{\text{(exp)}} = A_{\text{dn}}$ for monochromatic neutrinos and
\begin{equation}
    A_{\text{dn}}^{\text{(exp)}} = \frac{\int \rho(E)\sigma(\nu_e;E)\diff{E}  \times  A_{\text{dn}}(E) \; (1+T_{\text{MSW}}(E))}
                                        {\int \rho(E)\sigma(\nu_e;E)\diff{E}  \times (1+T_{\text{MSW}}(E)) } \qquad \text{(charged current only)}
\end{equation}
for continuous-spectrum neutrinos.

\subsection{The effect of the crust}\label{sec:CrustEffect}
The estimation \eqref{Adn} shown above is substantially based on the piecewise continuous structure
of the density profile and shows that the asymmetry should depend only on the density of rock
immediately under the detector, i.e. in the Earth's crust. At the same time, for beryllium
neutrinos, the actual width of the crust is comparable with the oscillation length, and neither
the valley nor the cliff approximation is valid for this layer. For boron neutrinos, the
crust can be considered a cliff, however, the oscillation length becomes comparable with the
thickness of the upper mantle, as well as with thicknesses of transition zones (see
Fig.~\ref{fig:EarthDensity} and \cite{Anderson1}). Moreover, the phase incursions are small within
the crust (and the upper mantle), and so are their time variations, hence, we are unable
to use the stationary phase approximation. However, we are still able to account for the effect of
these near-surface layers on the observed day-night asymmetry factor \eqref{Adn}, relying
upon relatively small density variation $\sDelta\eta$ within them. This feature, together
with the bounded layers' thickness, makes it possible to find the closed form of the
approximate flavor evolution operator for the crust (the upper mantle)
$[x_{n-1}^+,x_n^-]$ \ (see Appendix~\ref{app:crust}),
\begin{eqnarray}\label{eq:R0approx_crust}
    R_0(x_n^-, x_{n-1}^+) &=& \exp\{(-\ii\sigma_2\beta + \ii\sigma_1\alpha) e^{2\ii\sigma_3\psi(x_{n-1}^+)}\} + \bigO(\eta^2), \\
    \beta + \ii\alpha &=& \int\limits_{x_{n-1}^+}^{x_n^-}\dot{\theta}(y) e^{2\ii(\psi(y)-\psi(x_{n-1}^+))} \diff{y},
    \qquad \alpha,\beta = \bigO(\sDelta\eta) \in \mathds{R}.
\end{eqnarray}
This result formally repeats the cliff approximation \eqref{eq7} up to the substitution
$\sDelta\theta_j \to \beta$, $\mu_j \to \alpha$. Using such a substitution, the transition
zones, whose widths are comparable with the boron neutrino oscillation lengths, can be safely
replaced by the cliffs, preserving the form of expression \eqref{eq:T_approx}. Finally, to account
for the effect of the crust (and the upper mantle), as well as the effects of the transition zones,
we should make the following modification in \eqref{eq:T_approx}:
\begin{eqnarray}
    T_{\text{night}} &=& \cos2\theta_{\text{Sun}}
        \Bigl\{
                \cos2\theta_n^- +
                2\sin2\theta_n^- \sum_{j=1}^{n-1} (\sDelta\theta_j \cos2\sDelta\psi_{n,j} +
                                                   \bar\mu_j \sin2\sDelta\psi_{n,j})
    \nonumber\\
        &&\qquad\qquad - 4\sDelta\psi_{\text{det}} \sin2\theta_{\text{det}} \sum_{j=1}^{n-1}
        \sDelta\theta_j \sin2\sDelta\psi_{n,j}
        \Bigr\} + \sDelta T_{\text{night}},
    \label{eq:T_approx_crust}
    \\
    \sDelta T_{\text{night}} &=& \cos2\theta_{\text{Sun}} \{2\sin2\theta_n^- (\beta \cos2\sDelta\psi_{n,n-1} +
    \alpha \sin2\sDelta\psi_{n,n-1})
    \nonumber\\
    &&\qquad\qquad -4\sDelta\psi_{\text{det}}\sin2\theta_{\text{det}} (\beta\sin2\sDelta\psi_{n,n-1} - \alpha\cos2\sDelta\psi_{n,n-1})\}
    \nonumber\\
    &=& 2\cos2\theta_{\text{Sun}} \sin2\theta_n^-
    \int\limits_{x_{n-1}^+}^{x_n^-}\dot\theta(y) \cos2\sDelta\psi(y) \;\diff{y}
    - 4 \cos2\theta_{\text{Sun}} \sDelta\psi_{\text{det}} \sin2\theta_{\text{det}}
    \int\limits_{x_{n-1}^+}^{x_n^-}\dot\theta(y) \sin2\sDelta\psi(y) \;\diff{y} ,
\end{eqnarray}
where $\sDelta\psi(y) \equiv \psi(x_n^-) - \psi(y)$. The leading $\bigO(\eta)$ correction of the
crust (the upper mantle) to the day-night asymmetry factor \eqref{Adn} reads
\begin{equation}\label{Adn_crust}
    \sDelta A_{\text{dn}} =
               \frac{T_{\text{MSW}}}{1+T_{\text{MSW}}} \cdot \frac{\sin^2 2\theta_0}{\cos2\theta_0}\frac{2 E}{\sDelta m^2}
               \int\limits_{x_{n-1}^+}^{x_n^-}  \dot{V}(y) \cos2\sDelta\psi(y) \,\diff{y}.
\end{equation}
Again, this result should be subjected to the averaging procedure to acquire the predictive power.
However, due to the boundedness of the cosine, the correction is easily estimated (both before and
after averaging),
\begin{equation}\label{eq:Crust_uncert}
    \left|\frac{(\sDelta A_{\text{dn}})}{A_{\text{dn}}}\right| \le
    \frac{1}{V(x_n^-)}\int\limits_{x_{n-1}^+}^{x_n^-} |\dot{V}(y)|\diff{y} \sim
    \frac{|(\sDelta V)_{\text{crust}}|}{V(x_n^-)}.
\end{equation}

\section{Discussion}\label{sec:Discussion}
Let us begin with the review of the approximations we were using in the above calculations of the
day-night asymmetry. First, the Sun was considered one big valley, i.e. the adiabatic approximation
was used for it. The non-adiabatic corrections are of the order of the adiabaticity parameter
$\gamma = |\dot\theta(x)| / \lambda \omega$ \cite{MikheevSmirnov}. Using the fact that the typical
spatial scale of the solar density variation is associated with the solar core radius $R_0 \sim 0.1
R_{\text{Sun}} \approx 7\times10^4~\text{km}$ \cite{SolarDensity}, we find
\begin{equation}
     \gamma_{\text{Sun}} \lesssim
     \begin{cases}
         10^{-5},          &\quad E \sim 1~\text{MeV},\\
         5 \times 10^{-4}, &\quad E \sim 10~\text{MeV}.
     \end{cases}
\end{equation}
Thus, these corrections can be safely neglected.

Neglecting the finite width of the detector introduces a relative error of the order
$\bigO\left(\frac{L_{\text{det}}}{\ell_{\text{osc}}} \ \delta \right)$, compared with the leading
term \eqref{Adn}, as one can see from \eqref{eq:T_approx}. This correction is minuscule even for
beryllium neutrinos ($\ell_{\text{osc}} \approx 30~\text{km}$), since the detector sizes are now
$L_{\text{det}} \le 1~\text{km}$. Moreover, this type of correction is subjected to additional
suppression due to time averaging.

The valley-crust approximation considered in Sec.~\ref{sec:EvolutionOperator} relies upon
the smallness of parameters $\eta$ and $\delta$. The relative error of the linear approximation in
$\eta$ is of the order of $\bigO(\eta)$, hence, this approximation works fine for beryllium
neutrinos and quite well for boron neutrinos (see Eq.~\eqref{eq:eta_constraints}; note that the
maximum values specified for $\eta$ correspond to the inner core, while for typical neutrino
detector latitudes, the Sun never descends low enough to shine through it). Thus, for boron
neutrinos, the error of the linear approximation is within $10\%$. Further, the error due to the finite
width of the $j$th valley is estimated as $\bigO(\sDelta\eta_j \delta_j) = \bigO(\sDelta\eta_j \
\ell_{\text{osc}}/L_j)$, where $\sDelta\eta_j$ is the density variation on the $j$th valley and
$L_j$ is the width of the valley. On the other hand, if the width of the $k$th cliff $L_k \not\ll
\ell_{\text{osc}}$, the resulting error is $\bigO(\eta \delta_k) = \bigO(\eta \ L_k /
\ell_{\text{osc}})$. Within the PREM model, the cliffs are abrupt ($\delta_k = 0$), so we will not
discuss the numerical values of the corresponding errors.

If there are deep layers in the PREM density distribution, whose widths are of the order of the
oscillation length, they can be considered neither valleys nor cliffs. However, if the density
variations $\sDelta\eta$ over these layers are small, we can use the approximation considered in
Sec.~\ref{sec:CrustEffect} and Appendix~\ref{app:crust}, which formally reduces these layers to the
cliffs (see Eq.~\eqref{eq:R0approx_crust}). The error of such an approach is measured by the
magnitudes of parameters $\alpha, \beta$ in \eqref{eq:R0approx_crust}, i.e. the error is of the
order of $\sDelta\eta$. The contributions of such layers are additionally suppressed after
averaging, due to their depth (see below). In the case of the Earth, there is a layer of this type.
Namely, the core-mantle transition zone is about $200\,\text{km}$ wide \cite{Anderson1}, which is
of the order of the typical boron neutrino oscillation length. However, the corresponding density
variation lies within only $\sDelta\eta / \eta \sim 5-7\%$, and the time-averaged contribution is
obviously negligible. For beryllium neutrinos, the core-mantle transition zone can be considered a
valley.

The contributions from the Earth's layers, which lie much more than the oscillation length
under the detector, are suppressed due to time averaging, the principal nonvanishing contributions
being provided by the winter/summer solstice stationary points. These contributions have the
relative magnitude $\bigO\left(\frac{\sDelta\theta_j}{\sDelta\theta_n} \frac{r_j}{r_n}
\frac{\ell_{\text{osc}}}{L_{n,j}}\right)$ compared with the leading term \eqref{Adn} \ (see
Eqs.~\eqref{eq:T_approx} and \eqref{eq:YearAverage}). This suppression is quite strong for dense
though deep inner Earth's shells, including its core.

Finally, for the layers which lie within $\bigO(\ell_{\text{osc}})$ under the detector
and have the widths comparable with the oscillation length, the averaging procedure cannot be
performed so as to result in the expression \eqref{eq:YearAverage}. Here, we are only able to
follow the approach of Sec.~\eqref{sec:CrustEffect}, which results in an unaveraged correction
\eqref{Adn_crust}. The latter correction, in principle, can be explicitly averaged using numerical
integration, which in this case involves only the near-surface Earth's structure.
The strict constraint on the uncertainty introduced by neglecting the effect of this structure is
given in Eq.~\eqref{eq:Crust_uncert}. This expression predicts the $20\%$ uncertainty for beryllium
and $\sim30\%$ for boron neutrinos. However, it is quite obvious that the time average of the
cosine in \eqref{Adn_crust} is sensitive to the Wolfenstein potential in the layers within
$\bigO(\ell_{\text{osc}})$ under the detector; moreover, \eqref{eq:YearAverage} shows that this
sensitivity asymptotically falls off as the inverse depth of the layer $L_{n,j}$. Thus, if the
resulting expression \eqref{Adn} is used for boron neutrinos, one should substitute for $V(x_n^-)$
the Wolfenstein potential averaged over the crust and the uppermost layers of the mantle, using
some decreasing weighting function. The resulting potential will be $15-20\%$ larger than that
immediately under the Earth's surface. For beryllium neutrinos, it suffices to substitute the
Wolfenstein potential in the crust.

Thus, we may conclude that, according to \eqref{Adn} and \eqref{Adn_crust}, the asymmetry
has the order $\bigO(\eta)$ and is determined by the rock density in the layer of width about
$\ell_{\text{osc}}$ under the detector, i.e. in the Earth's crust (as well as in the uppermost part
of the mantle for boron neutrinos). The principal uncertainty of expression \eqref{Adn} comes from
the fact that the stationary point approximation we used for the analytical time averaging of
Eq.~\eqref{eq:T_approx} is inapplicable to the layers which lie within several neutrino oscillation
lengths under the detector. Another correction comes from the time averaging of oscillating
contributions of the deep layers; using \eqref{eq:YearAverage}, we can estimate its magnitude as
$\le 3\%$ for beryllium neutrinos and $\le 10-15\%$ for boron neutrinos with $E=10~\text{MeV}$. All
other corrections, taken together, do not exceed $10\%$.


\vspace{0.5em}

Using the recent data from SNO, KamLAND,  and Borexino collaborations \cite{SNO, KamLand,
Borexino1}, namely, $\tan^2 \theta_0\approx 0.46$ and $\sDelta m^2\approx 7.6\times
10^{-5}~\text{eV}^2$, and the typical electron densities in the Earth's crust $N_{e(\text{crust})}=
1.3\,\text{mol}/\text{cm}^3$ \cite{Anderson1} and in the solar core $N_{e(\text{Sun})} \sim
100~\text{mol}/\text{cm}^3$ \cite{SolarDensity}, we arrive at the numerical estimation for the
day-night asymmetry factor for solar beryllium-7 neutrinos ($E = 0.862\,\text{MeV}$)
\begin{equation}\label{eqf}
   A_{\text{dn}}({}^7\text{Be}) = (-4.0 \pm 0.9) \times 10^{-4}, \qquad A_{\text{dn}}^{\text{(exp)}}({}^7\text{Be}) = (-2.6 \pm 0.6) \times 10^{-4}.
\end{equation}
The uncertainty corresponds to the effect of the Earth's crust, which, as mentioned, can be
explicitly evaluated by numerical averaging of Eq.~\eqref{Adn_crust}. For boron-8
neutrinos with $E = 10~\text{MeV}$, substitution of the crust density into \eqref{Adn} leads to the
asymmetry estimation
\begin{equation}\label{eqf_boron}
    A_{\text{dn}}({}^8\text{B}) \approx (2.9 \pm 0.8)\times 10^{-2} \qquad (E = 10 \text{~MeV}).
\end{equation}
The `experimental' asymmetry factor for the electron scattering experiment with the recoil
kinetic energy threshold $T_{\text{min}} = 4.5\text{~MeV}$ (such as Super-Kamiokande-III \cite{SuperKIII_2010}),
averaged over boron-8 solar neutrino spectrum \cite{Boron8Spectrum}, is then given by Eq.~\eqref{Adn_exp_continuous},
\begin{equation}\label{eqf_boron_exp}
    A_{\text{dn}}^{\text{(exp)}}({}^8\text{B}) \approx (1.6 \pm 0.5)\times 10^{-2} \qquad \text{(averaged over energy)},
\end{equation}
the fully-numerical calculation yielding $A_{\text{dn}}^{\text{(exp)}}({}^8\text{B}) =
1.9\times10^{-2}$. We do not present here the asymmetry predictions for ${}^{13}\text{N}$,
${}^{15}\text{O}$, and $pep$ neutrinos, since their typical energies are around $1\text{~MeV}$,
while the fluxes at least one order smaller than that for beryllium neutrinos \cite{SolarDensity}.

\begin{figure}[h]
  \includegraphics[width=14cm]{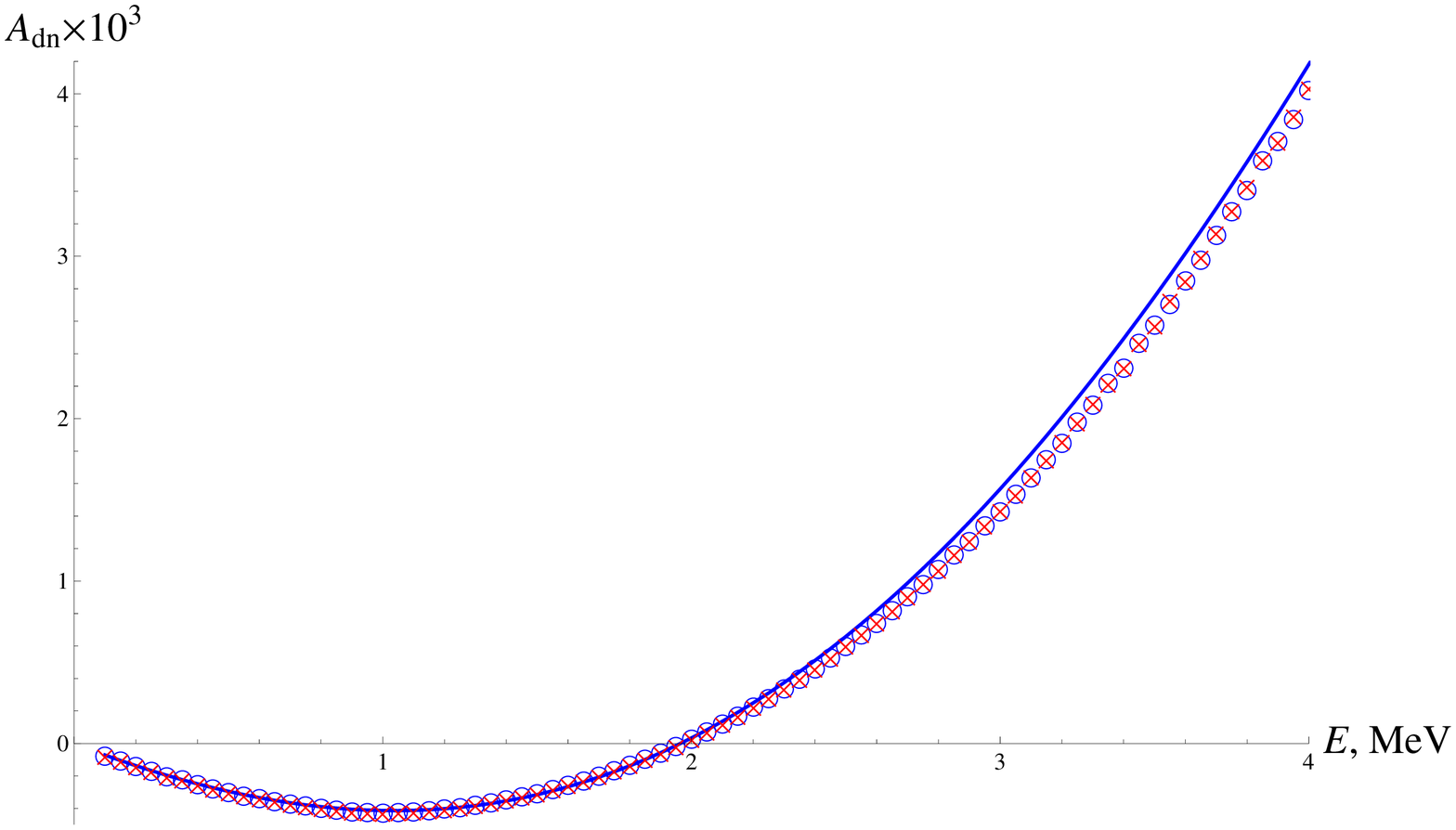} \\
  \text{(a)}\\[0.5em]
  \includegraphics[width=9cm]{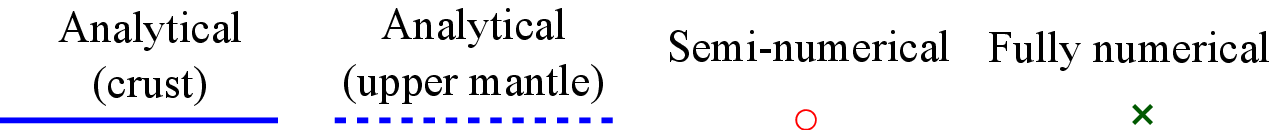}\\[0.5em]
  \includegraphics[width=14cm]{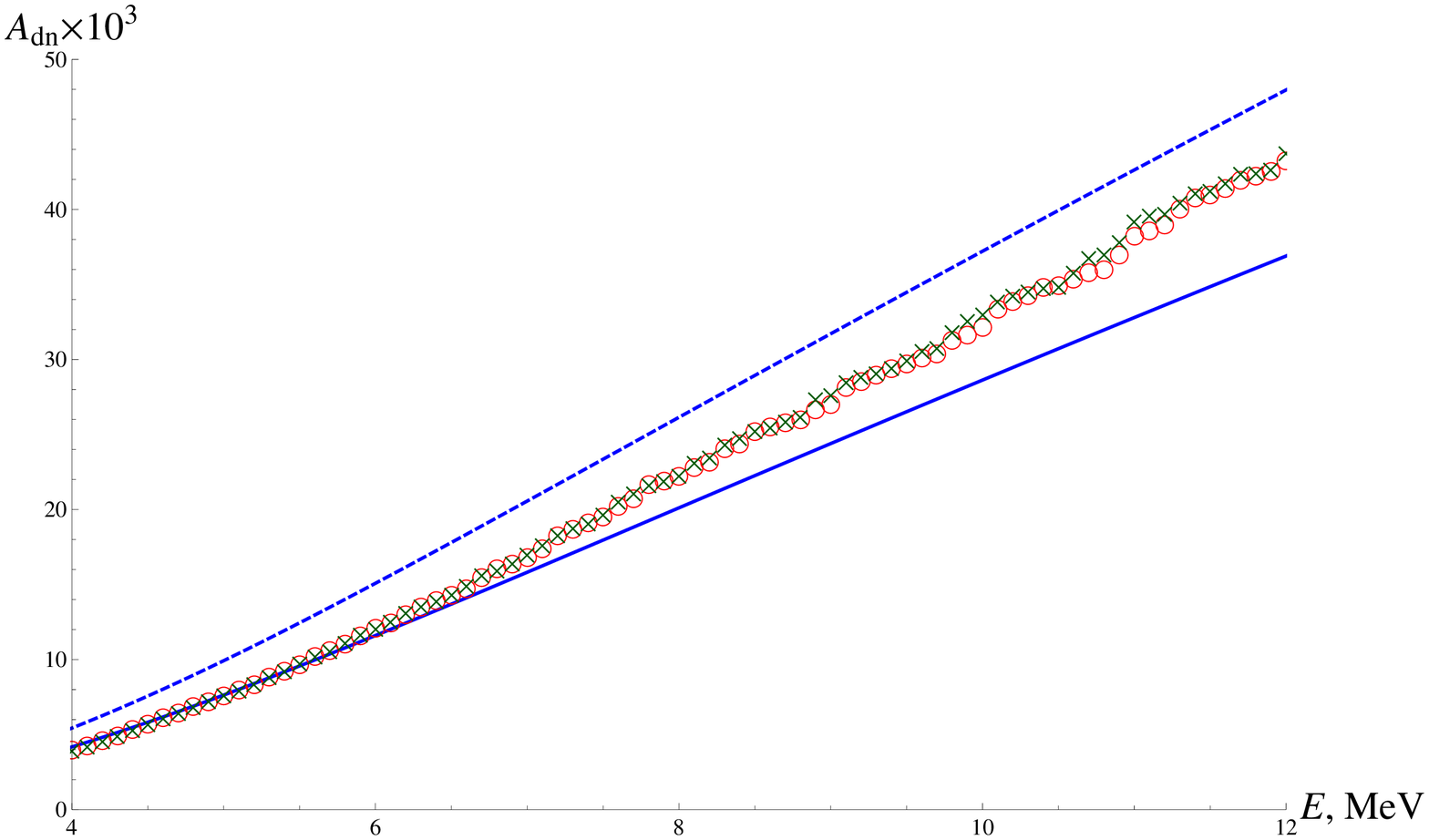} \\
  \text{(b)}
  \caption{Comparison of analytical expression \eqref{Adn} for the day-night asymmetry factor with
  the results of numerical simulations for Gran Sasso ($\chi = +42.5\degree$).
  Solid/dashed curves: Analytical estimations \eqref{Adn} with the crust/upper mantle electron
  densities substituted; Circles ($\bigcirc$): The result based on numerical averaging of analytical expression \eqref{eq:T_approx} (`valley-cliff'
  approximation); Crosses ($\times$): Fully numerical result (numerical solution of \eqref{eq4} and subsequent numerical averaging).
  Subfigures (a) and (b) demonstrate the low- and high-energy segments of the curves, respectively. }
  \label{fig:Adn_E}
\end{figure}

If, for boron neutrinos, one substitutes into \eqref{Adn} the mean density in the near-surface layers,
instead of the crust density, the above predictions \eqref{eqf_boron}, \eqref{eqf_boron_exp}
will be about 10\% larger (depending on the weighting function chosen for the mean density evaluation)
but never larger than those obtained by substituting the density
of the upper mantle. Indeed, Fig.~\ref{fig:Adn_E}b shows that the two analytical curves
corresponding to the two `surface densities' discussed here enclose the fits obtained using
two techniques of numerical simulation. One of these simulations involves the numerical time averaging of the leading
$\bigO(\eta)$ terms in expression \eqref{eq:T_approx} (the valley-cliff approximation), while the
other one includes both the numerical solution of the evolution equation \eqref{eq4} and the
subsequent time averaging.

\begin{figure}[h]
  \includegraphics[width=16cm]{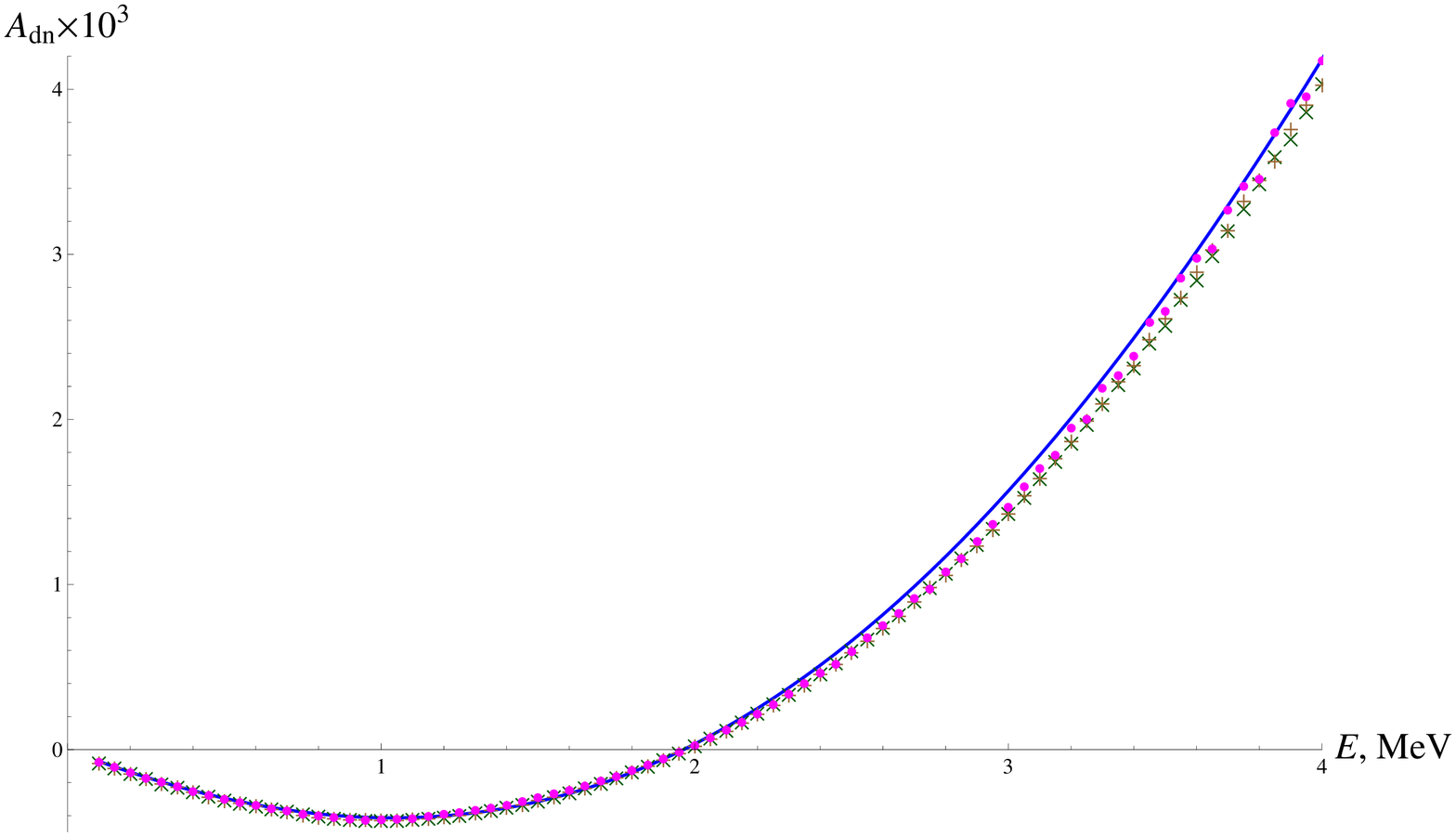} \\
  \text{(a)} \\[0.5em]
  \includegraphics[width=8.5cm]{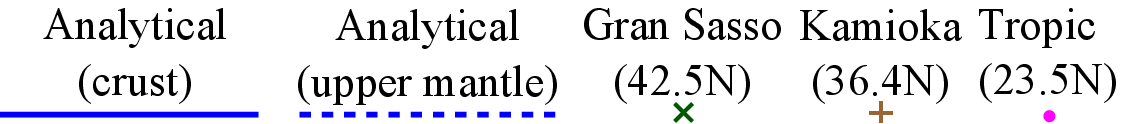}\\[0.5em]
  \includegraphics[width=16cm]{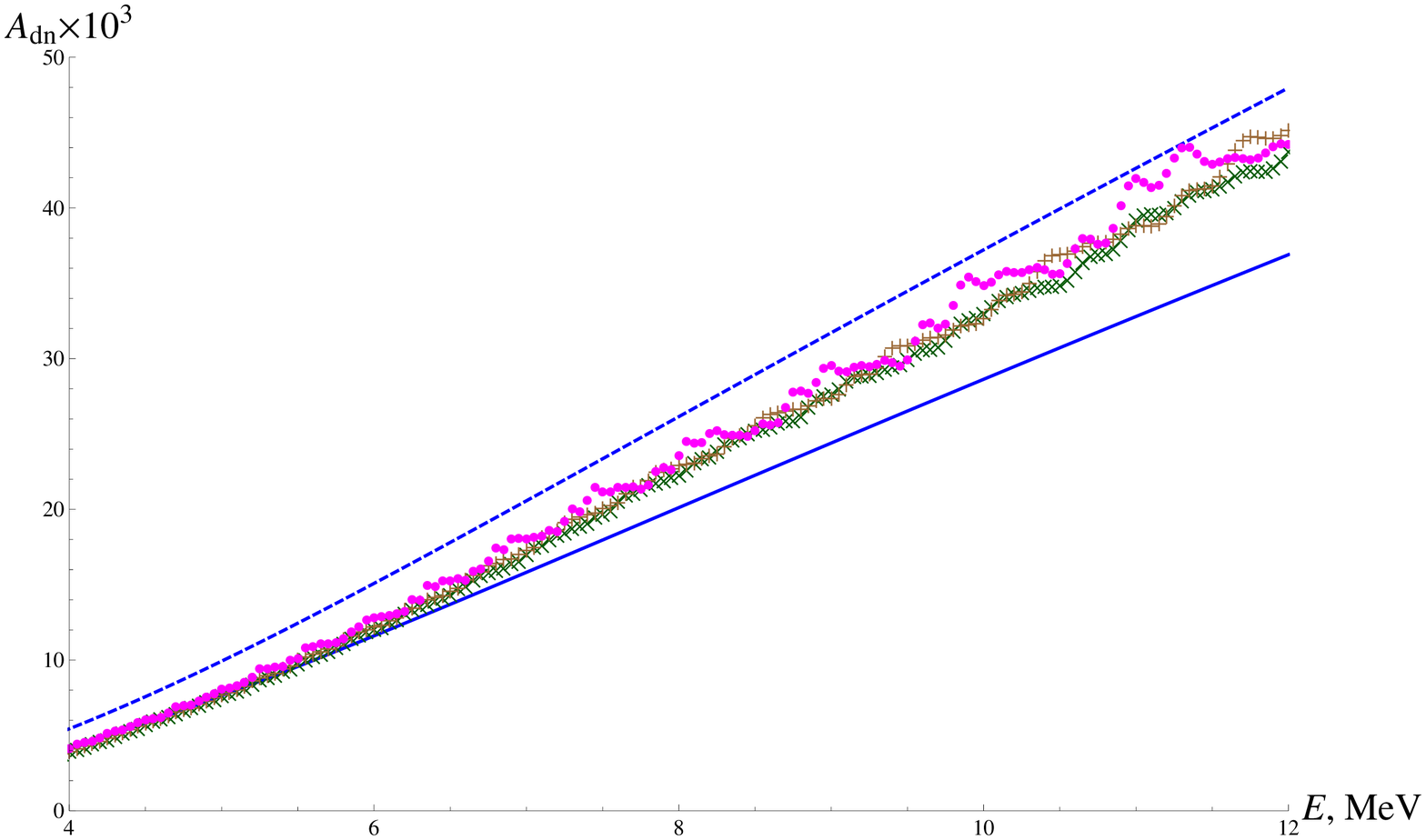} \\
  \text{(b)}
  \caption{Comparison of analytical expression \eqref{Adn} for the day-night asymmetry factor with
  the results of numerical simulations for different detector latitudes.
  Solid/dashed curves: Analytical estimations \eqref{Adn} with the crust/upper mantle electron
  densities substituted; Crosses ($\times$): numerical result for Gran Sasso ($\chi =
  +42.5\degree$), Pluses ($+$): for Kamioka ($\chi = +36.4\degree$), Bullets ($\bullet$): for the Northern
  tropic ($\chi = +23.5\degree$).
  Subfigures (a) and (b) demonstrate the low- and high-energy segments of the curves, respectively.
  }
  \label{fig:Adn_E_latitudes}
\end{figure}

The numerical curves shown in Fig.~\ref{fig:Adn_E} were calculated for the Gran Sasso
laboratory, where the Borexino detector is operating; the curves for Kamioka (Super-Kamiokande) and
Sudbury (SNO) lie very close to that for Gran Sasso, so we do not show them in this figure. Instead, a
comparison of numerical results for selected detector latitudes is presented in
Fig.~\ref{fig:Adn_E_latitudes} (again, we do not include the SNO latitude, since the results for it lie very close to those for Gran Sasso).
We have also included in Fig.~\ref{fig:Adn_E_latitudes} the numerical results for the Northern tropic, since,
according to our analytical estimations, near it,
the subleading contributions to the asymmetry may become considerable, which come from the solstices (see Eq.~\eqref{eq:YearAverage}).
We will address this issue further in this section.

From Fig.~\ref{fig:Adn_E_latitudes}, one may infer that the leading-order analytical estimation \eqref{Adn} is in good agreement with the numerical results even for boron neutrinos.
Contrary to these numerical results, however, our analytical estimation is model-independent,
in particular, it does not contain the latitude of the neutrino detector. The possible errors of our predictions can also be
easily estimated. In the case of beryllium neutrinos, the estimation of the errors becomes strict
enough to result in a fixed-boundary interval (not a confidence interval) for the day-night
asymmetry shown in Eq.~\eqref{eqf}, which is useful for experimental purposes. Our results are also
in agreement with numerical day-night asymmetry predictions presented by other authors (see, e.g.,
\cite{BahcallKrastev}).

As predicted, the agreement of the numerical results with the analytical one becomes better for
low-energy neutrinos. We also observe that the asymmetry vanishes for low-energy neutrinos, which
is a result of the applicability of the adiabatic approximation for such neutrinos. Indeed, within
this approximation, the neutrino flavor observation probabilities depend only on the creation and
absorption points.

It is worth emphasizing here that such a directly measurable quantity, as the day-night asymmetry
factor, does depend both on the neutrino regeneration effect inside the Earth and on the
Mikheev--Smirnov--Wolfenstein effect inside the solar core, where the neutrino is created. Another
useful physical quantity, namely, the regeneration factor $f_{\text{reg}}$, describes solely the
Earth effect in the day-night asymmetry \cite{Wei},
\begin{eqnarray}\label{f_reg}
    A_{\text{dn}} &=& -\frac{2\cos{2\theta_\text{Sun}}}{1 + (T_{\text{night}}+T_{\text{day}})/2}
    f_{\text{reg}} \approx -\frac{2\cos{2\theta_\text{Sun}}}{1 + T_{\text{MSW}}}
    f_{\text{reg}}, \\
    f_{\text{reg}} &\approx& \frac12 \sin^2 2\theta_0 \frac{2 E V(x_n^-)}{\sDelta m^2}. \label{f_reg_value}
\end{eqnarray}
Indeed, one observes that, unlike the regeneration factor \eqref{f_reg_value}, the day-night asymmetry \eqref{Adn} depends on
the solar effect manifested in the energy-dependent quantity $T_{\text{MSW}}$. As a result, at
the energy $E \sim 2.0\,\text{MeV}$, which corresponds to the Mikheev--Smirnov resonance in the
solar core, we have $\cos2\theta_{\text{Sun}} = 0$ and $T_{\text{MSW}} = 0$, and, as a consequence, the
day-night asymmetry factor \eqref{Adn} changes sign (see Fig.~\ref{fig:Adn_E}a); on the other hand,
the regeneration factor is always positive. Thus, it is the resonance inside the Sun which makes the asymmetry,
observed on the Earth, vanish.

From Fig.~\ref{fig:Adn_E}a,~\ref{fig:Adn_E_latitudes}a, one may observe that beryllium neutrinos
($E = 0.862\,\text{MeV}$) would be indeed quite useful for the study of the matter effects in the
neutrino oscillations, since they correspond to almost maximum asymmetry magnitude in the unusual
domain of its negativity. Moreover, solar beryllium neutrinos are highly monochromatic (in
contrast, e.g., to the boron neutrinos) and their flux is considerably larger \cite{SolarDensity}.
However, we are able to conclude that the day-night effect needs at least a 10--20 times
improvement of current experimental resolution to be distinguished at a considerable confidence
level for such neutrinos. In particular, the result of the Borexino experiment
$A^{(\text{exp})}_{\text{dn}}({}^7\text{Be}) = \bigl(1 \pm 12 \text{(stat.)}\pm 7
\text{(syst.)}\bigr)\times 10^{-3}$ was reported in April, 2011 \cite{Borexino} and demonstrates
the strong dominance of the uncertainties over the expected effect. One may still hope that the
50~kton LENA detector which should come into operation around 2020 \cite{LENA} and is expected to
observe as much as $10^4$ solar beryllium neutrino events per day, could distinguish the day-night
effect for such neutrinos. Strictly speaking, the Poisson statistics results in the relative errors
$\Delta N / N \sim 1/\sqrt{N}$, and, for a year-long experiment, one reaches the statistical error
of the order of $1/\sqrt{365\cdot10^4} \sim 0.5\times10^{-3}$. However, one could employ the
adaptive processing of the experimental data, taking into account the expected form of the curve
$T_{\text{night}}(\Theta_{\text{N}})$ (see \eqref{eq:T_approx} and \eqref{eq:Lnj_ThetaZ}), i.e. the
dependence of the asymmetry on the nadir angle. This processing technique may be efficient in
extracting the day-night effect from under the noise even for small event rates. Quite a similar
technique was recently suggested in \cite{LENA_FluxVariations} as a search tool for periodic time
variations in the ${}^7\text{Be}$ solar neutrino flux observed at LENA.

In Fig.~\ref{fig:Adn_E_latitudes}b, one can also observe the role of the stationary phase
points in the time average of the day-night asymmetry factor. Namely, the numerical curve
corresponding to the Northern tropic demonstrates specific oscillations which come from the
amplified contribution of the winter solstice stationary point to the year average of the asymmetry
factor (see Eq.~\eqref{eq:YearAverage}). The curve for the Kamioka mine, which is situated about
1.5 times closer to the Tropic than Gran Sasso, also demonstrates oscillations, compared to the
Gran Sasso curve. In view of this effect, it would be quite prospective to build a detector
close the Tropic, which could be able to observe high-energy solar neutrinos with an energy
resolution about $0.5~\text{MeV}$. A favorable place for such a high-technology project could be,
e.g., near S\~ao Paulo, Brazil (latitude $\chi = -23\degree33'$, i.e. exactly on the Southern tropic!),
especially under the potential support of the local university.

One should hold in mind here that the positions of the interference peaks on Fig.~\ref{fig:Adn_E_latitudes} 
substantially depend on the radii of the Earth's shells. Nevertheless, the approximate smoothness of the three numerical curves in
Fig.~\ref{fig:Adn_E_latitudes} 
indicates the (approximate) stability of the day-night asymmetry
factor with respect to slight variations of the parameters of the PREM model, namely, the radii of
the Earth's shells and the density jumps. As seen from Fig.~\ref{fig:Adn_E_latitudes}, this stability becomes stronger for low-energy neutrinos,
as well as for the detectors operating far from the Tropic.

\vspace{1em}

It is also interesting to study the effect of the local Earth's crust inhomogeneities under the
detector on the observed day-night asymmetry. Such inhomogeneities could be associated, for
instance, with oil-bearing horizons. Let the inhomogeneity be described by the variation $\delta
N_e^{\text{(i)}}(x)$ of the electron density over the smooth profile $\bar{N}_e(x)$,
\begin{equation}
    N_e(x) = \bar{N}_e(x) + \delta N^{\text{(i)}}_e(x), \qquad \delta N^{\text{(i)}}_e(x) = 0 \text{ for } x \not\in [x_{\text{i}}, x_{\text{i}} + \delta
    x_{\text{i}}],
\end{equation}
where the inhomogeneity size $\delta x_{\text{i}} \ll \ell_{\text{osc}}$. Then the contribution of
this inhomogeneity to the asymmetry factor is given by Eq.~\eqref{Adn_crust},
\begin{eqnarray}
    \frac{\delta A_{\text{dn}}^{\text{(i)}}}{A_{\text{dn}}} &=& -\frac{1}{N_{e\text{(crust)}}}
               \int\limits_{x_{\text{i}}}^{x_{\text{i}} + \delta x_{\text{i}}}  \delta\dot{N}_{e}^{\text{(i)}}(y) \cos2\sDelta\psi(y)
               \diff{y} = \frac{1}{N_{e\text{(crust)}}}
               \int\limits_{x_{\text{i}}}^{x_{\text{i}} + \delta x_{\text{i}}}  \delta
               N_{e}^{\text{(i)}}(y)\sin2\sDelta\psi(y)
               \frac{2\pi\omega(y)\diff{y}}{\ell_{\text{osc}}},\\
    \left|\frac{\delta A_{\text{dn}}^{\text{(i)}}}{A_{\text{dn}}}\right| &\le& \frac{|\delta
    N_e^{\text{(i)}}|}{N_{e\text{(crust)}}} \frac{2\pi\delta x_{\text{i}}}{\ell_{\text{osc}}}\,
    \left|\sin\frac{2\pi L^{\text{(i)}}}{\ell_{\text{osc}}}\right|,
\end{eqnarray}
where $L^{\text{(i)}} = x_n^- - x_{\text{i}}$ is the depth of the inhomogeneity under the detector.
One can see that, although the regeneration effect is stronger for higher-energy
neutrinos, this effect is insensitive to the near-surface local inhomogeneities for such neutrinos
due to the smallness of the sine and ratio $2\pi\delta x_{\text{i}}/\ell_{\text{osc}}$ in the above
expression. Therefore, exploration of the Earth's crust based on the neutrino oscillations would
require an extreme improvement of current measurement techniques \cite{OilDetection}, and its
prospects seem obscure in the nearest future.

\section{Conclusion}\label{sec:Conclusion}%

Let us make a brief summary of our initial goals concerning the analytical approach to the
day-night asymmetry and the results of our investigation. We have attempted to develop a framework
able to give interval constraints on the day-night flavor asymmetry, which are independent of the
details of the density distribution inside the Earth. Of course, some approximations should be made
to analytically obtain such general results; in our case, the principal assumptions were the
relatively small density of the Earth ($\eta \ll 1$) and its spherically-symmetric layered
structure (manifested in the small parameter $\delta$). Although the actual approximation
parameters may be not quite small, using the framework developed, we can readily estimate the
corresponding inaccuracies in our predictions (as done, e.g., in Sec.~\ref{sec:Discussion}).

Our analysis shows that the day-night asymmetry is insensitive to the structure of the deep Earth's
layers, including its core; we found that this sensitivity falls off as the inverse layer's depth
$1/L_{n,j}$ (see \eqref{eq:YearAverage}). The day-night effect averaged over time (i.e. the effect
directly measured in neutrino experiments) is principally determined by the mean electron density
of the Earth within $1-2$ oscillation lengths under the neutrino detector, i.e. in the Earth's
crust for beryllium neutrinos, as well as in the upper mantle for boron neutrinos. The
corresponding leading-order analytical curves plotted for these two electron densities are shown in
Fig.~\ref{fig:Adn_E} and~\ref{fig:Adn_E_latitudes} (the solid and the dashed line, respectively),
together with the results of the numerical simulations. One may see that the leading approximation
works  quite well within the energy range $E \sim 0.5-12~\text{MeV}$ under major interest in the
field. Moreover, in Fig.~\ref{fig:Adn_E}b, one may notice that high-energy neutrinos, as it was
expected, `sense' the deeper Earth's layers. Indeed, it is indicated by the fact that the
high-energy segment of the numerical curve approaches the dashed theoretical curve plotted for the
(higher) upper mantle density.
If, in view of this fact, one substitutes into the leading-order expression \eqref{Adn} the mean
Earth's density within $1-2$ oscillation lengths under the detector, the resulting estimation will
agree with the numerical one within $10\%$. Such an accuracy of the day-night effect measurements
is yet to be achieved in the future experiments, such as LENA \cite{LENA}.


Further, the theoretical analysis predicts next-to-leading-order corrections to the day-night
effect, which may arise due to the stationary phase points during the year, at which the nighttime
neutrino oscillation phase freezes. These points occur on the winter and the summer solstices and
are spectacular for the fact that their contributions do not vanish after (arbitrarily) long
observations. These contributions are very small for low-energy neutrinos, as well as in the
temperate latitudes, however, in the tropical latitudes, they are quite distinguishable. One can
see these stationary point contributions in Fig.~\ref{fig:Adn_E_latitudes}b (high-energy segment of
Fig.~\ref{fig:Adn_E_latitudes}), especially for the Tropical curve (latitude $\chi = +23.5\degree$)
demonstrating oscillations relative to the other curves plotted for the more temperate
latitudes.
With the present energy resolutions reaching $0.5~\text{MeV}$ \cite{SuperKIII_2010},
only the event rates are yet too small to observe this effect.

It is also worth mentioning that the good agreement of our analytical predictions with the
numerical simulations in a surprisingly wide range of neutrino energies (i.e. oscillation lengths)
is also a byproduct of a number of specific features of the actual Earth's density profile. For
instance, the large density jumps are lying very deep inside the Earth; the crust is quite thin
(and, as mentioned, is not a cliff for beryllium neutrinos), however, its density is quite low; the
layers' widths are incomparable, etc. The framework developed in the present paper, however, is
able to reveal the situations in which the `universality' of the prediction \eqref{Adn} will not
hold; one needs only the general features of the density distribution to make such a conclusion.

\section*{Acknowledgments}
The authors are grateful to A.~V.~Borisov and V.~Ch.~Zhukovsky for fruitful discussions of the
ideas of the present paper. The authors would also like to thank Wei Liao for his stimulating
remarks concerning the averaging procedure. Finally, we should thank E.~Lisi and D.~Montanino for
Fig.~1 in their paper \cite{Lisi}, which we have used for creating Fig.~\ref{fig:EarthDensity} in
the present manuscript. The numerical simulations reported in the manuscript were made using the
Supercomputing cluster ``Lomonosov'' at the Moscow State University \cite{LomonosovCluster}.

\appendix
\section{Approximate solutions for the evolution operator}
\label{app:solutions}
In this section, we derive the approximate solutions for the flavor evolution operator $R_0(x,x_0)$
(see Eqs.~\eqref{eq2} and \eqref{eq4}) for valleys and cliffs. Our approximations will
only deal with the neutrino propagation inside the Earth, since the neutrino propagation inside the
Sun is highly adiabatic (see the adiabaticity estimations in Sec.~\ref{sec:Discussion}), i.e.,
$R_0 = 1$ with a great accuracy. Moreover, the Earth regeneration effect under investigation depends only on the Earth's density
distribution $N_e(x)$ and not on the details of the neutrino propagation inside the Sun.

Now, due to the relatively small Earth's density, which manifests itself as a small parameter $\eta
\lesssim 10^{-1}$, we resort to the linear approximation in $\eta$, namely, take the two leading
terms of the Dyson series
\begin{equation}\label{Texp_expansion}
    R_0(x,x_0) = \TProd \exp \left\{ -\ii\sigma_2 \int\limits_{x_0}^x \dot\theta(y) e^{2\ii\sigma_3\psi(y)} \diff{y}
    \right\} = \idMatrix -\ii\sigma_2 \int\limits_{x_0}^x \dot\theta(y) e^{2\ii\sigma_3\psi(y)}
    \diff{y} + \bigO(\eta^2).
\end{equation}

\subsection{Valleys} In the valleys, we have a smooth and bounded function $\dot\theta(y)$ and a
rapidly oscillating matrix exponential $e^{2\ii\sigma_3\psi(y)} = \cos2\psi(y) +
\ii\sigma_3\sin2\psi(y)$. Then, applying the double integration by parts, we arrive at
\begin{eqnarray}\label{valley_int_byParts}
    \int\limits_{x_0}^{x} \dot{\theta}(y) e^{2\ii\sigma_3\psi(y)} \diff{y} &=&
    \left.\frac{(1-\mathcal{D}_y)\dot\theta(y)}{2\ii\sigma_3\lambda \omega(y)} e^{2\ii\sigma_3\psi(y)}
    \right|_{x_0}^x + \int\limits_{x_0}^{x} \mathcal{D}_y^2 \dot{\theta}(y) \cdot e^{2\ii\sigma_3\psi(y)}
    \diff{y},\\
    \mathcal{D}_y f(y) &\equiv& \frac{\pd}{\pd y} \left(\frac{1}{2\ii\sigma_3\lambda\omega(y)}
    f(y)\right).
\end{eqnarray}
Let $\sDelta\eta$ be the total variation of the density parameter $\eta$ on the valley.
Then the variation of the effective mixing angle $\theta(x)$ has the order $\bigO(\sDelta\eta)$,
and all the gradients in the above expressions are suppressed by powers of the small parameter
$\delta = \ell_{\text{osc}} / L = \pi / \lambda L$, where $L$ is the width of the valley and
$\ell_{\text{osc}}$ is the oscillation length. Namely,
\begin{eqnarray}
  \dot\theta &=& \bigO(\sDelta\theta / L) = \bigO(\sDelta\eta/L), \\
  \ddot\theta &=& \bigO(\sDelta\theta / L^2) = \bigO(\sDelta\eta/L^2),  \\
  \omega &=& \sqrt{a^2 + b^2} = 1 + \bigO(\eta),          \label{omega_estim} \\
  \dot\omega &=& \frac{2 a \omega}{b}\dot\theta = \bigO(\sDelta\eta / L).
\end{eqnarray}
Using these estimations, one can readily show that
\begin{eqnarray}
    \mathcal{D}_y \dot\theta(y) &=& \frac{1}{2\ii\sigma_3\lambda}\left( \frac{\ddot{\theta}}{\omega} - \frac{\dot\omega \dot\theta}{\omega^2}
    \right) = \bigO\left(\frac{\sDelta\eta}{\lambda L^2}\right) + \bigO\left(\frac{(\sDelta\eta)^2}{\lambda L^2}\right) = \bigO(\sDelta\eta \ \delta/L),\\
    \mathcal{D}_y^2 \dot\theta(y) &=& \bigO(\sDelta\eta \ \delta^2/L),
\end{eqnarray}
then the matrix norm of the remainder integral in Eq.~\eqref{valley_int_byParts}
\begin{equation}
    \left\| \int\limits_{x_0}^{x} \mathcal{D}_y^2 \dot{\theta}(y) \cdot e^{2\ii\sigma_3\psi(y)}
    \diff{y} \right\| \le \int\limits_{x_0}^{x} \| \mathcal{D}_y^2 \dot{\theta}(y) \| \diff{y} =
    \bigO(\sDelta\eta \ \delta^2).
\end{equation}
On the other hand, the first term in the right side of Eq.~\eqref{valley_int_byParts} consists of
two parts, the one proportional to $\mathcal{D}_y\dot\theta(y)$, which is of the order
$\bigO(\sDelta\eta \ \delta / \lambda L) = \bigO(\sDelta\eta \ \delta^2)$, and the other
proportional to $\dot\theta(y)$, which is $\bigO(\sDelta\eta \ \delta)$. Then, summarizing the
estimations made, we conclude that
\begin{equation}
    \int\limits_{x_0}^{x} \dot{\theta}(y) e^{2\ii\sigma_3\psi(y)} \diff{y}
    = \left.\frac{\dot\theta(y)}{2\ii\sigma_3\lambda \omega(y)}
    e^{2\ii\sigma_3\psi(y)}\right|_{x_0}^x + \bigO(\sDelta\eta \ \delta^2).
\end{equation}
Moreover, here, due to \eqref{omega_estim}, we can safely replace  $\omega(y)$ by unity, and then
the approximate solution of the evolution equation in the valley takes the form
\begin{equation}
   R_0(x,x_0) = \idMatrix - \frac{\ii\sigma_1}{2\lambda}\left(\dot\theta(x) e^{2\ii\sigma_3\psi(x)} - \dot\theta(x_0)
   e^{2\ii\sigma_3\psi(x_0)}\right) + \bigO(\sDelta\eta \ \delta^2).
\end{equation}
Finally, by neglecting the terms of the order $\bigO((\sDelta\eta)^2\delta^2)$, we can write
\begin{equation}\label{R0approx_valley}
   R_0(x,x_0) = \exp\left\{-\frac{\ii\sigma_1}{2\lambda}\left(\dot\theta(x) e^{2\ii\sigma_3\psi(x)} - \dot\theta(x_0)
   e^{2\ii\sigma_3\psi(x_0)}\right)\right\} + \bigO(\sDelta\eta \ \delta^2).
\end{equation}
This expression can also be derived using mathematically strict stationary phase technique
\cite{StationaryPhase}.

\subsection{Cliffs}
On the cliffs, the change of the effective mixing angle is of the order $\bigO(\eta)$, while the
spatial scale of this change is quite small. Therefore, the change of the phase of oscillations
$\sDelta\psi \ll 2\pi$, and we can expand the exponential in the right side of
Eq.~\eqref{Texp_expansion} in the local phase incursion
\begin{eqnarray}
   \psi(y) - \psi(x_0) &=& \lambda \int\limits_{x_0}^{y} \omega(z)\diff{z} = \lambda(y-x_0) + \bigO(\eta\delta),\\
   e^{2\ii\sigma_3 \psi(y)} &=& e^{2\ii\sigma_3\psi(x_0)} (1 + 2\ii\sigma_3\lambda (y-x_0)) + \bigO(\delta^2) + \bigO(\eta\delta).
\end{eqnarray}
Now, using the boundedness of the total variation of the mixing angle on the cliff, we obtain
\begin{equation}
   \int\limits_{x_0}^x \dot\theta(y) e^{2\ii\sigma_3\psi(y)}\diff{y} = e^{2\ii\sigma_3\psi(x_0)} \int
   \limits_{x_0}^x (1 + 2\ii\sigma_3\lambda (y-x_0))\dot\theta(y)\diff{y} + \bigO(\eta\delta^2).
\end{equation}
This finally leads to the cliff approximation for the evolution operator
\begin{eqnarray}
    R_0(x,x_0) &=& \idMatrix
               + (-\ii\sigma_2 \sDelta\theta + \ii\sigma_1\mu)e^{2\ii\sigma_3\psi(x_0)}
               + \bigO(\eta\delta^2)
    \nonumber\\
               &=& \exp\{(-\ii\sigma_2\sDelta\theta + \ii\sigma_1 \mu)e^{2\ii\sigma_3\psi(x_0)}\} + \bigO(\eta\delta^2) + \bigO(\eta^2),
    \label{R0approx_cliff}
\end{eqnarray}
where $\sDelta\theta \equiv \theta(x) - \theta(x_0) = \bigO(\eta)$ and
\begin{equation}\label{mu_def}
   \mu = 2\lambda \int\limits_{x_0}^x (y-x_0) \dot\theta(y)\diff{y} = \bigO(\eta\delta).
\end{equation}

\subsection{Valley + cliff}\label{app:valley_cliff}
It is also useful to consider a valley $[x_{j}^+, x_{j+1}^-]$ following a cliff $[x_{j}^-,
x_{j}^+]$. Using expressions \eqref{R0approx_valley} and \eqref{R0approx_cliff} for the evolution
operators on these segments, within the linear accuracy in $\eta$, we obtain
\begin{gather}
    R_0(x_{j+1}^-, x_{j}^-) = R_0(x_{j+1}^-, x_{j}^+) R_0(x_{j}^+, x_{j}^-)
    \nonumber\\
    = \exp\left\{
             - \frac{\ii\sigma_1}{2\lambda}\left(\dot\theta(x_{j+1}^-) e^{2\ii\sigma_3\psi(x_{j+1}^-)} - \dot\theta(x_{j}^+) e^{2\ii\sigma_3\psi(x_{j}^+)}\right)
             + (-\ii\sigma_2\sDelta\theta_{j} + \ii\sigma_1\mu_{j})e^{2\ii\sigma_3\psi(x_{j}^-)}
          \right\} + \bigO(\eta\delta^2)
    \nonumber\\
    = \exp\left\{
             -\frac{\ii\sigma_1}{2\lambda}\dot\theta(x_{j+1}^-) e^{2\ii\sigma_3\psi(x_{j+1}^-)}
             +\left(-\ii\sigma_2\sDelta\theta_{j}
             + \ii\sigma_1\bar\mu_{j}
             \right)e^{2\ii\sigma_3\psi(x_{j}^-)}
          \right\} + \bigO(\eta \delta^2),
\end{gather}
where $\sDelta\theta_{j} = \theta(x_{j}^+) - \theta(x_{j}^-)$, $\bar\mu_{j} = \mu_{j} +
\dot\theta(x_{j}^+) / 2\lambda$, and $\mu_j$ is defined analogously to \eqref{mu_def}. Now, by
substituting the above result into representation \eqref{eq2} for the evolution operation, we
arrive at
\begin{gather}
    R(x_{j+1}^-, x_{j}^-) = e^{\ii\sigma_2\theta(x_{j+1}^-)} e^{\ii\sigma_3\psi(x_{j+1}^-)}
                            R_0(x_{j+1}^-, x_{j}^+) R_0(x_{j}^+, x_{j}^-)
                            e^{-\ii\sigma_3\psi(x_{j}^-)}e^{-\ii\sigma_2\theta(x_{j}^-)}
    \nonumber\\
    = e^{\ii\sigma_2\theta(x_{j+1}^-)}\left[
             e^{\ii\sigma_3\sDelta\psi_{j+1}} - \frac{\ii\sigma_1}{2\lambda}\dot\theta(x_{j+1}^-) e^{\ii\sigma_3\sDelta\psi_{j+1}}
             +\left(-\ii\sigma_2\sDelta\theta_{j}
             + \ii\sigma_1\bar\mu_{j}
             \right)e^{-\ii\sigma_3\sDelta\psi_{j+1}}
      \right]e^{-\ii\sigma_2\theta(x_{j}^-)}+ \bigO(\eta\delta^2)
    \nonumber\\
    = e^{\ii\sigma_2\theta(x_{j+1}^-)}e^{-\ii\sigma_1\dot\theta(x_{j+1}^-)/2\lambda} e^{\ii\sigma_3\sDelta\psi_{j+1}}
    e^{\ii\sigma_1\bar\mu_{j}} e^{-\ii\sigma_2\sDelta\theta_{j}}e^{-\ii\sigma_2\theta(x_{j}^-)} + \bigO(\eta\delta^2) +
    \bigO(\eta^2), \label{R_valley_cliff}
\end{gather}
where $\sDelta\psi_{j+1} \equiv \psi(x_{j+1}^-) - \psi(x_{j}^-)$. This expression, in turn, can be
generalized to a sequence of $n$ valleys (see Eq.~\eqref{eq9}).

\subsection{The Earth's crust}\label{app:crust}
As one could see from the main flow of the paper, for neutrinos with energies $E \sim
1-10~\text{MeV}$, the Earth's crust (and, for $E\sim 10~\text{MeV}$, the upper mantle) cannot be
considered either a valley or a cliff, since its width is comparable with the oscillation length,
and the parameter $\delta$ is of the order of unity. Here, however, another approximation is
useful, which takes into account the small density variation $\sDelta\theta$ over these
layers, as well as their bounded thickness. In terms of parameters $\eta$ and $\delta$, we have
\begin{eqnarray}
    \eta = \frac{2 E V}{\sDelta m^2} &\lesssim&
                            \begin{cases}
                                 2 \times 10^{-3} &\quad\text{(beryllium neutrinos, $E = 0.862$~MeV, the crust),} \\
                                 3 \times 10^{-2} &\quad \text{(boron neutrinos,  $E = 10$~MeV, crust + upper mantle),}
                            \end{cases}
    \\
    \frac{\sDelta\eta}{\eta} &\le& 0.3 \quad \text{\;(both cases)},\\
    \delta = \frac{L}{\ell_{\text{osc}}} &\lesssim& 5 \qquad \text{(both cases)}.
\end{eqnarray}
Then, using expansion \eqref{Texp_expansion}, we find the immediate expression for the evolution
operator for the crust:
\begin{equation}\label{R0approx_crust}
    R_0(x,x_0) = \idMatrix + (\ii\sigma_1 \alpha - \ii\sigma_2 \beta)e^{2\ii\sigma_3\psi(x_0)} + \bigO(\eta^2) =
    \exp\{(-\ii\sigma_2\beta+\ii\sigma_1\alpha) e^{2\ii\sigma_3\psi(x_0)}\} + \bigO(\eta^2),
\end{equation}
where real numbers $\alpha, \beta = \bigO(\sDelta\eta)$ are defined by the expression
\begin{equation}
    \beta + \ii\alpha = \int\limits_{x_0}^{x} \dot{\theta}(y) e^{2\ii(\psi(y)-\psi(x_0))} \diff{y}.
\end{equation}
Note that the form of this approximation \eqref{R0approx_crust} coincides with the cliff
approximation \eqref{R0approx_cliff}, up to the coefficient substitution $\sDelta\theta \to \beta$,
$\mu \to \alpha$. In particular, the cliff approximation is restored in the $\delta \to 0$ limit.

\end{document}